\documentclass[%
 reprint,
 amsmath,amssymb,
 aps,
prc,
]{revtex4-2}

\usepackage{amsmath,amssymb}
\usepackage{bm}
\usepackage{graphicx}
\usepackage{dcolumn}
\usepackage{tabularx}
\usepackage{enumitem}
\usepackage{threeparttable}
\usepackage{booktabs}
\usepackage{cancel}
\usepackage[normalem]{ulem}
\usepackage{xcolor}
\usepackage{physics}
\usepackage{float}

\usepackage[colorlinks=true,citecolor=blue,urlcolor=blue,linkcolor=blue]{hyperref}

\begin{document}

\title{Neutron star matter with hyperons: Bayesian comparison of nucleonic and SU(6)/SU(3) hyperonic models}


%\author{$^{1}$}
 %\altaffiliation[Also at ]{Indian Institute of Technology Jodhpur, Jodhpur 342037 India}%Lines break automatically or can be forced with \\
% \email{ }
%\author{ $^{2}$}
%\author{ $^{3}$}
%\author{ $^{1, 4}$}
%\affiliation{$^{1}$ }%
%\affiliation{$^{2}$ }
%\affiliation{$^{3}$ }
%\affiliation{$^{4}$}

\author{Athira S.$^{1}$}
\author{Vishal Parmar$^{2}$}
\email{vishal.parmar@pi.infn.it}
\author{Monika Sinha$^{1}$}
\author{Ignazio Bombaci$^{2,3}$}

\affiliation{$^{1}$ Indian Institute of Technology Jodhpur, Jodhpur 342037, India}
\affiliation{$^{2}$ INFN, Sezione di Pisa, Largo B. Pontecorvo 3, I-56127 Pisa, Italy}
\affiliation{$^{3}$ Dipartimento di Fisica, Universit\`{a} di Pisa, Largo B. Pontecorvo 3, I-56127 Pisa, Italy}


\begin{abstract}
We investigate neutron star matter with hyperons within a density-dependent relativistic mean-field framework using Bayesian inference, considering three composition scenarios: purely nucleonic matter, hyperonic matter under SU(6) flavor symmetry, and hyperonic matter under SU(3) symmetry with free vector-sector parameters.
%For the first time, the full spectrum of hyperonic degrees of freedom is consistently explored within the SU(3) framework. 
The analysis incorporates constraints from empirical nuclear matter properties, theoretical inputs at low densities, and multimessenger observations of neutron stars. We find that the SU(6) scheme, grounded in the quark model and isospin counting rule, leads to a significantly softer equation of state. In contrast, the additional flexibility of the SU(3) framework enhances vector repulsion and yields a comparatively stiffer equation of state consistent with observational bounds across the explored parameter space; in particular, the posterior distributions favor values of the vector coupling ratio $\alpha_v$ lower than the SU(6) limit $\alpha_v = 1$. These differences are reflected in neutron star observables, including mass--radius relations, tidal deformabilities, direct Urca thresholds, and oscillation properties, all of which remain compatible with current constraints within the SU(3) scenario. We further examine structural signatures through the curvature of the mass--radius relation and find that, although hyperon-rich configurations can induce noticeable variations, such features depend sensitively on the stiffness of the equation of state and are therefore not universally robust indicators. Bayesian model comparison further shows that present constraints do not meaningfully discriminate between the purely nucleonic and SU(3) hyperonic scenarios, while providing positive, but not decisive, evidence against the more restrictive SU(6) framework.
\end{abstract}

\maketitle

%\tableofcontents

\section{Introduction}
\label{introduction}

A major uncertainty in the physics of neutron star (NS) interiors concerns the role of hyperons in dense baryonic matter. Their appearance is energetically favored at high densities, but the resulting changes in the equation of state (EOS) depend sensitively on the poorly constrained strong interactions in the strange baryon sector. More broadly, despite extensive studies, the exact composition of NS interiors remains uncertain. Various possibilities have been proposed, including the presence of heavy baryons \cite{2020Parti...3..660T, 2023PrPNP.13104041S,   2025ApJ...980...54L}, hyperons \cite{Vida_a_2016, Zachariou:2024apf, Weissenborn:2011kb, PhysRevC.81.035803, Logoteta:2021iuy, Schaffner:1995th}, meson condensates \cite{2024PhRvC.110d5804P,2001PhRvC..64e5805B}, and deconfined quark matter \cite{1999PhRvC..60a5802P, 1992ApJ...400..647G, 1996PhR...264..143G, Lugones_PRD_2005, 2011PhRvD..83b5012L,2010PhRvD..82f5017L, 2011JKPS...59.2114H}. Once exotic degrees of freedom appear, baryonic matter fragments into multiple Fermi components, altering the EOS and its response to compression, with direct implications for transport properties and oscillation modes \cite{Zachariou:2024apf, PhysRevC.109.L032801, Gaitanos_2021}. In particular, while hyperon formation is energetically favored at high densities, their presence generally softens the EOS, making it difficult to sustain the large masses observed in NSs, a tension known as the hyperon puzzle \cite{Bombaci2021_Hyp-puzzle, Vidana:2010ip, PhysRevC.87.055801, PhysRevLett.114.092301, Tolos:2017lgv, PhysRevC.61.055801, Logoteta_2019}.   

Importantly, this is not simply a numerical mismatch but reflects a deeper uncertainty in how particles interact in dense matter. The degree to which the EOS softens depends strongly on the interplay between attractive scalar interactions and repulsive vector interactions in the strange sector. Additional effects, such as repulsive channels, many-body contributions \cite{PhysRevC.95.044308, PhysRevC.90.045805}, momentum-dependent interactions \cite{Sun:2022yor,v21s-d9dt, PhysRevC.111.054605, FOLIAS2025122982,Chen:2013uua}, and hyperonic three-body interactions of the type \(NNY\), \(NYY\), and \(YYY\), can partially counterbalance the softening \cite{PhysRevLett.114.092301,Logoteta_2019}. In particular, repulsive \(YY\) channels mediated by hidden-strangeness vector mesons, density-dependent couplings, higher-order mesonic terms in relativistic mean-field models, and chiral \(NN\Lambda\) three-body forces have all been shown to delay hyperon onset and stiffen the EOS \cite{Logoteta_2019, PhysRevC.90.045805}. Recent \textit{ab initio} calculations have further emphasized this point by including \(N\Lambda\), \(\Lambda\Lambda\), \(NN\Lambda\), and \(N\Lambda\Lambda\) interactions and showing that hyperonic matter can remain compatible with present mass, radius, and tidal-deformability constraints \cite{PhysRevLett.114.092301, Tong_2025}. In this sense, the hyperon puzzle should be viewed not as a contradiction but as a useful constraint on the nature of dense matter interactions, bridging nuclear physics experiments with astrophysical inference \cite{Haskell_2010, Jyothilakshmi_2022, Nayyar_2006, Malik:2025exa}. 

The nuclear physics input to the strange sector is informed by \(\Lambda\) hypernuclear spectroscopy, where high-resolution experiments, particularly at JLab, probe the \(\Lambda N\) interaction and constrain hyperonic interactions relevant for dense matter \cite{Hashimoto:2006aw, Gal_2016}. Astrophysical inference complements this by constraining dense matter through observations of NS masses, radii, and tidal deformabilities from gravitational waves, along with cooling and timing measurements, thereby placing empirical bounds on the EOS and the possible presence of strangeness. Representative high-precision mass measurements further reinforce these constraints, including PSR J1614$-$2230 (\(M = 1.97 \pm 0.04\, M_\odot\)) \cite{2010ApJ...724L.199O}, PSR J0740+6620 (\(M = 2.08 \pm 0.07\, M_\odot\) with 95\% credibility) \cite{2021ApJ...918L..28M}, PSR J0348+0432 (\(M = 2.01 \pm 0.04\, M_\odot\)) \cite{Bandyopadhyay:2021gnp,2018ChJPh..56..292H}, PSR J1810+1744 (\(M = 2.13 \pm 0.04\, M_\odot\)) \cite{2021ApJ...908L..46R}, and PSR J0952$-$0607 (\(M = 2.35 \pm 0.17\, M_\odot\)) \cite{2022ApJ...934L..17R}, which impose stringent lower bounds on the maximum mass, thereby disfavoring excessively soft EOSs and constraining the role of exotic degrees of freedom such as hyperons.

In hyperonic dense matter, symmetry principles play a central role in constraining the otherwise poorly known interactions in the strange sector. In particular, the choice between SU(6) spin--flavor symmetry and the more general SU(3) flavor symmetry leads to qualitatively different prescriptions for hyperon--meson couplings, thereby affecting the composition and stiffness of the EOS, with direct consequences for NS and observables. While SU(6) symmetry provides a commonly adopted and more restrictive framework, SU(3) symmetry allows greater flexibility in the vector interaction sector, enabling a broader exploration of dense-matter properties \cite{PhysRevC.81.025801, PhysRevD.107.036011, huang2022hyperonicstarrelativisticmeanfield}. Given the limited experimental constraints on hyperonic interactions, it becomes essential to systematically test how these symmetry assumptions translate into observable consequences in dense matter.

Motivated by this interplay, the inclusion of hyperons in NS matter, first proposed by Ambartsumyan and Saakyan \cite{Ambartsumyan_Saakyan}, has been extensively studied using microscopic \cite{PhysRevLett.114.092301, PhysRevC.90.045805, Katayama:2015dga, PhysRevC.84.035801} and phenomenological approaches \cite{2014PhLB..734..383V, PhysRevC.87.055806, PhysRevC.85.015804, Maslov_2015, Oertel_2015}, including relativistic mean-field models with density-dependent and SU(3)/SU(6)-motivated couplings \cite{ Bednarek_2012, PhysRevC.95.065803,1963RvMP...35..916D}, microscopic Brueckner-Hartree-Fock calculations \cite{PhysRevC.81.035803, Katayama:2015dga, Miyatsu_2015, PhysRevC.58.3688,katayama2014}, Skyrme-type energy-density functionals \cite{Lim_2015, Mornas:2004vs}, chiral and \textit{ab initio} approaches including hyperonic three-body forces \cite{Logoteta_2019, Tong_2025, Petschauer_2017}, and momentum-dependent or non linear derivative models \cite{chorozidou2024, KOCHANKOVSKI2026140129}, with recent efforts focusing on constraining the EOS through astrophysical observations using statistical inference and machine-learning techniques \cite{PhysRevD.98.023019, PhysRevD.101.054016, Ferreira:2019bny, Morawski:2020izm}, and on constructing hyperonic EOSs consistent with massive pulsars while identifying robust multimessenger and statistical signatures of hyperons \cite{Zachariou:2024apf, Miyatsu:2025rzn, bauswein2025, PhysRevD.110.123016, PhysRevD.106.063024}. The main challenge is the large degeneracy in the hyperonic parameter space. Hypernuclear data provide only partial constraints, while symmetry prescriptions such as SU(6) and SU(3) restrict the couplings without determining them uniquely. As a result, different descriptions of the strange sector can produce EOSs consistent with observations but with very different internal compositions \cite{PhysRevD.106.063024, Logoteta:2021iuy}. This ambiguity motivates a statistical framework that can systematically explore the allowed models.

In this work, we address this using a density-dependent relativistic mean-field (DDRMF) framework with explicit inclusion of hyperons, treating the underlying couplings as parameters to be inferred within a unified Bayesian approach. We consider three scenarios: purely nucleonic matter, hyperonic matter under SU(6) symmetry, and hyperonic matter within SU(3) symmetry with free vector-sector parameters, while constraining scalar couplings using hypernuclear data. Here, we explore the full SU(3) parameter space through posterior sampling. In contrast to studies that consider a more limited subset of hyperons \cite{PhysRevD.106.063024}, we include the full hyperon spectrum, allowing for a broader and more flexible characterization of NS observables, such as masses, radii, and tidal deformabilities, and offering complementary insights into the dense-matter EOS alongside existing approaches \cite{PhysRevC.98.035804, PhysRevD.100.023015, 10.1093/mnras/stw2152}. It is important to emphasize that consistency with current neutron-star observations does not by itself constitute a resolution of the hyperon puzzle. A genuine microscopic resolution requires hyperonic interactions that are simultaneously constrained by hypernuclear data and derived or validated within a controlled many-body framework. In the present work, the SU(6) and SU(3) prescriptions are instead used as phenomenological parametrizations of the hyperon--meson couplings. Our aim is therefore to determine which regions of the adopted parameter space remain compatible with current nuclear and multimessenger constraints. The central question addressed here is therefore not whether hyperons must be present in neutron-star cores, but whether current nuclear and multimessenger data can distinguish hyperonic matter from nucleonic matter once the uncertain strange-sector interaction is allowed to vary.

The paper is structured as follows. Section \ref{formalism} provides a concise overview of the DDRMF formalism utilized in this study. We present the priors and constraints for our Bayesian analysis. Section \ref{Results} presents the results, wherein we examine the posterior distribution of several properties of nuclear matter and NSs, along with their associated interactions. In Sec. \ref{Summary}, we provide a summary of our work.

\section{Formalism}
\label{formalism}
\subsection{RMF model for dense matter}
The dense matter considered in this work consists of baryon octet $(n,\, p,\, \Lambda,\, \Sigma^{+},\, \Sigma^{0},\, \Sigma^{-},\, \Xi^{0},\, \Xi^{-})$ in $\beta$ equilibrium with leptons $(e^-, \mu^-)$. The system is described within the relativistic mean-field (RMF) framework \cite{walecka1974theory, Parmar_2021}, where the strong interactions among hadrons are mediated by the isoscalar–scalar $\sigma$, isoscalar–vector $\omega^{\mu}$ and $\phi^{\mu}$, and isovector–vector $\rho^{\mu}$ meson fields. Accordingly, the Lagrangian density is written as \cite{Schaffner:1995th, Schaffner:1993nn}
\begin{equation}
\begin{aligned}
\mathcal{L}_B 
&= \sum_B \bar{\psi}_B 
\Big[ 
    i\gamma_\mu \partial^\mu - m_B 
    + g_{\sigma B}\sigma 
    - g_{\omega B}\gamma_\mu \omega^\mu 
    - g_{\phi B}\gamma_\mu \phi^\mu \\
&\quad - g_{\rho B}\gamma_\mu \vec{\tau}_B \cdot \vec{\rho}^\mu \Big] \psi_B + \frac{1}{2}\left( 
    \partial_\mu \sigma \, \partial^\mu \sigma 
    - m_\sigma^2 \sigma^2 
\right) - U(\sigma) \\
&\quad - \frac{1}{4} \omega_{\mu\nu} \omega^{\mu\nu} 
+ \frac{1}{2} m_\omega^2 \omega_\mu \omega^\mu - \frac{1}{4} \vec{\rho}_{\mu\nu} \cdot \vec{\rho}^{\mu\nu} 
+ \frac{1}{2} m_\rho^2 \vec{\rho}_\mu \cdot \vec{\rho}^\mu \\
&\quad
- \frac{1}{4} \phi_{\mu\nu} \phi^{\mu\nu} 
+ \frac{1}{2} m_\phi^2 \phi_\mu \phi^\mu .
\end{aligned}
\end{equation}

The baryonic degrees of freedom are described by the Dirac spinor $\psi_B$, with the vacuum mass $m_B$ and the isospin operator $\vec{\tau}_B$. The quantities $\omega_{\mu\nu}$, $\phi_{\mu\nu}$, and $\rho_{\mu\nu}$ correspond to the field strength tensors of the $\omega$, $\phi$, and $\rho$ mesons, respectively. The scalar self-interaction term \cite{Boguta:1977xi} is expressed as
\begin{equation}
U(\sigma) = \frac{1}{3} g_2 \sigma^3 + \frac{1}{4} g_3 \sigma^4 .
\end{equation}
The calculation is carried out within the mean-field approximation \cite{Serot:1984ey}. The symbols $\sigma$, $\omega_0$, $\rho_{03}$, and $\phi_0$ represent the mean values of the respective meson fields. Consequently, the meson operators are replaced by their expectation values, leading to the meson-field equations in the form presented below.

\begin{eqnarray}
m_\sigma^2 \sigma = \frac{-\partial U}{\partial \sigma} +  \sum_B g_{\sigma B}n_B^s, \\ 
m_\omega^2 \omega_0 = \sum_B g_{\omega B}n_B, \\
m_\phi^2 \phi_0 = \sum_B g_{\phi B}n_B, \\
m_\rho^2 \rho_{03} = \sum_B g_{\rho B}I_{3B}n_B. 
\end{eqnarray}
The baryon scalar and vector number densities are given as
\begin{eqnarray}
n_{S}^B = \frac{2J_B + 1}{2\pi^2} \int_0^{k_{F_B}} \frac{m_B^{*}}{\sqrt{k^2 + m_B^{*2}}} k^2 dk, \\
n_B = (2J_B + 1) \frac{k_{F_B}^3}{6\pi^2},
\end{eqnarray}
where $J_B$, $k_{F_B}$ and $I_{3B}$ are the spin, Fermi momentum and the third component of the isospin projection.\\
The effective mass of a baryon $B$ is given by
\begin{equation}
m_B^{*} = m_B - g_{\sigma B}\,\sigma,
\end{equation}
and the corresponding chemical potential is
\begin{equation}
\mu_B = \sqrt{k_{F_B}^2 + m_B^{*2}}
+ g_{\omega B}\,\omega_0 
+ g_{\phi B}\,\phi_0 
+ I_{3B}\, g_{\rho B}\,\rho_{03}.
\end{equation}
Since the hadronic phase is required to satisfy charge neutrality, the total charge density is given by
\begin{equation}
Q = \sum_B q_B n_B - n_e - n_\mu = 0 ,
\end{equation}
with $q_B$ being the electric charge of baryon $B$, and $n_e$, $n_\mu$ the number densities 
of electrons and muons, respectively.
The chemical equilibrium condition for each species is 
\begin{equation}
    \mu_i=b_i\mu_n - q_i\mu_e,
\end{equation}
We denote the chemical potentials of the $i$th baryon, neutron, and electron by $\mu_i$, $\mu_n$, and $\mu_e$, respectively. The corresponding conserved charges, baryon number and electric charge, are given by $b_i$ and $q_i$.

The total energy density is given by  
\begin{equation}
\begin{aligned}
\varepsilon &= 
\tfrac{1}{2} m_{\sigma}^{2} \sigma^{2} 
+ \tfrac{1}{3} g_{2} \sigma^{3} 
+ \tfrac{1}{4} g_{3} \sigma^{4} 
+ \tfrac{1}{2} m_{\omega}^{2} \omega_{0}^{2} 
+ \tfrac{1}{2} m_{\rho}^{2} \rho_{03}^{2} \\
&\quad + \tfrac{1}{2} m_{\phi}^{2} \phi_{0}^{2} 
+ \sum_{B} \frac{2 J_{B} + 1}{2\pi^{2}} 
\int_{0}^{k_{F_{B}}} \sqrt{k^{2} + {m_{B}^{*}}^{2}} \, k^{2} \, dk \\
&\quad + \sum_{l} \frac{1}{\pi^{2}} 
\int_{0}^{k_{F_{l}}} \sqrt{k^{2} + m_{l}^{2}} \, k^{2} \, dk \, ,
\end{aligned}
\end{equation}
and the pressure is given by 

\begin{equation}
\begin{aligned}
P &= 
- \tfrac{1}{2} m_{\sigma}^{2} \sigma^{2} 
- \tfrac{1}{3} g_{2} \sigma^{3} 
- \tfrac{1}{4} g_{3} \sigma^{4} 
+ \tfrac{1}{2} m_{\omega}^{2} \omega_{0}^{2} 
+ \tfrac{1}{2} m_{\rho}^{2} \rho_{03}^{2} \\
&\quad + \tfrac{1}{2} m_{\phi}^{2} \phi_{0}^{2} 
+ \sum_{B} \frac{2 J_{B} + 1}{6 \pi^{2}}
\int_{0}^{k_{F_{B}}} \frac{k^{4}}{\sqrt{k^{2} + {m_{B}^{*}}^{2}}} \, dk\\
&\quad
+ \sum_{l} \frac{1}{3 \pi^{2}}
\int_{0}^{k_{F_{l}}} \frac{k^{4}}{\sqrt{k^{2} + m_{l}^{2}}} \, dk \, .
\end{aligned}
\end{equation}

\subsection{Vector meson-hyperon couplings from SU(6) flavor symmetry}
In the SU(6) spin–flavor symmetry framework, a unified depiction of strongly interacting matter at high densities is attained by incorporating hyperons into the baryonic sector alongside nucleons. The study determines the baryon–meson coupling constants utilizing the quark model and isospin counting (QMIC) rule \cite{athira2025kaon}, which are given as follows
\begin{align}
    \frac{1}{2} g_{\omega \Lambda} &= \frac{1}{2} g_{\omega \Sigma} = g_{\omega \Xi }= \frac{1}{3} g_{\omega N} ,\\
    2 g_{\phi \Lambda} &= 2 g_{\phi \Sigma} = g_{\phi \Xi}= -\frac{2\sqrt{2}}{3} g_{\omega N} ,\\
   g_{\rho \Lambda} &= 0 ,\\
    \frac{1}{2} g_{\rho \Sigma} &= g_{\rho \Xi }= g_{\rho N}. 
\end{align}

\subsection{Vector meson-hyperon couplings from SU(3) flavor symmetry}
The framework of SU(3) flavor symmetry, which incorporates the up, down, and strange quarks, provides a natural and consistent basis for describing strongly interacting matter at high densities.  This method extends the baryonic sector to incorporate hyperons as inherent degrees of freedom in addition to nucleons. 

An invariant Yukawa Lagrangian can be formulated to describe the interactions of the baryon octet with the mediator nonet mesons \cite{1963RvMP...35..916D, 2014PhRvC..89b5805L},
\begin{equation}
\mathcal{L}_{\text{YUK}} = -\, g\, (\bar{\psi}_B \psi_B)\, M \, , \end{equation}
where $\psi_B$ denotes the Dirac field of the interacting baryons and $M$ represents the mediator meson field.
This Lagrangian belongs to the irreducible representation $\mathrm{IR}\{1\}$, corresponding to a unitary singlet. 
The vector mesons $\omega_8$ and $\rho_0$ belong to the vector octet $\mathrm{IR}\{8\}$, while $\phi_1$ is a vector singlet. To preserve unitary symmetry, the bilinear $(\bar{\psi}_B \psi_B)$ must transform as $\mathrm{IR}\{8\}$ when $M \in \mathrm{IR}\{8\}$ and as $\mathrm{IR}\{1\}$ when $M \in \mathrm{IR}\{1\}$.

Using the symmetric and antisymmetric couplings of $\mathrm{IR}\{8\}\otimes \mathrm{IR}\{8\} \to \mathrm{IR}\{8\}$, the Yukawa interaction can be written as
\begin{equation}
\mathcal{L}_{\text{YUK}} = - (g_{s} C_1 + g_{a} C_2 )\, (\bar{\psi}_B \psi_B) M 
\label{A2}
\end{equation}
for octet mediator mesons, and

\begin{equation}
\mathcal{L}_{\text{YUK}} = - g_1M ,
\end{equation}

for singlet mediator mesons, where $g_1$ is the meson singlet coupling, $C_1$ and $C_2$ denote the SU(3) Clebsch–Gordan coefficients associated with the symmetric and antisymmetric couplings, respectively, and $g_s$ and $g_a$ are the symmetric and antisymmetric couplings.
 
We use the CG coefficients of Ref.~\cite{1964RvMP...36.1005M} and, following Ref.~\cite{1963RvMP...35..916D}, define the constants

\begin{equation}
g_8 = \sqrt{\frac{30}{40}}\, g_{s} + \sqrt{\frac{6}{24}}\, g_{a},
\qquad
\alpha_v = {\sqrt{\frac{6}{24}}\, \frac{g_{a}}{g_8}}.
\end{equation}
The flavor SU(3) symmetry involves three parameters: the weight factor $\alpha_v$ [$\alpha_v = F/(F+D)$, F- antisymmetric coupling and D- symmetric coupling], the coupling ratio $z = g_8/g_1$ where $g_8$ and $g_1$ is the octet and singlet coupling respectively; and the mixing angle $\theta_v$ \cite{athira2025kaon, 1979PhRvD..20.1633N, 1999PhRvC..59...21R}.
Hence, in terms of these parameters, we can write the baryon vector meson coupling constants as

\begin{align}
g_{\rho N} &= g_{8} , \nonumber \\
g_{\rho\Sigma} &= 2 g_{8} \alpha_v, \qquad
g_{\rho \Xi} = - g_{8} (1 - 2 \alpha_v), \nonumber \\
g_{\rho\Lambda} &= 0, \qquad
g_{\omega_8 N} = \frac{1}{\sqrt{3}} g_{8} (4 \alpha_v - 1), \nonumber \\
g_{\omega_8 \Sigma} &= \frac{2}{\sqrt{3}} g_{8} (1 - \alpha_v), \qquad
g_{\omega_8 \Xi} = - \frac{1}{\sqrt{3}} g_{8} (1 + 2 \alpha_v), \nonumber \\
g_{\omega_8 \Lambda} &= - \frac{2}{\sqrt{3}} g_{8} (1 - \alpha_v), \nonumber \\
g_{\phi_1 N} &= g_{\phi_1\Sigma} = g_{\phi_1\Lambda } = g_{\phi_1\Xi} = g_{1}.
\end{align}

In nature, the physical isospin-singlet vector mesons $\omega$ and $\phi$ arise as mixtures of the idealized $\omega_8$ and $\phi_1$ states~\cite{1984PrPNP..12..171D}:
\begin{align}
\omega &= \cos\theta_v \ket{\omega_8} + \sin\theta_v \ket{\phi_1}, \nonumber \\
\phi   &= -\sin\theta_v \ket{\omega_8} + \cos\theta_v \ket{\phi_1}.
\end{align}

Here $\theta_v$ denotes the mixing angle between the $\omega_8$ and $\phi_1$ states. The couplings of the physical $\omega$ meson are then given by
\begin{align}
g_{\omega N} &= \cos\theta_v \, g_{1}
+ \sin\theta_v \, \frac{1}{\sqrt{3}} \, g_{8} \, (4\alpha_v - 1) \\ \nonumber
g_{\omega\Lambda} &= \cos\theta_v \, g_{1}
- \sin\theta_v \, \frac{2}{\sqrt{3}} \, g_{8} \, (1 - \alpha_v)  \\\nonumber
g_{\omega\Sigma} &= \cos\theta_v \, g_{1}
+ \sin\theta_v \, \frac{2}{\sqrt{3}} \, g_{8} \, (1 - \alpha_v) \\\nonumber
g_{\omega\Xi} &= \cos\theta_v \, g_{1}
- \sin\theta_v \, \frac{1}{\sqrt{3}} \, g_{8} \, (1 + 2\alpha_v).
\end{align}
The $\phi$-meson couplings follow from the $\omega$-meson couplings under the substitutions $\sin\theta_v \to \cos\theta_v$ and $\cos\theta_v \to -\sin\theta_v$~\cite{1984PrPNP..12..171D}.
From the above expressions, considering an ideal mixing angle ($\tan {\theta_v}= \frac{1}{\sqrt{2}}$), the meson-hyperon couplings can be written as follows: \\
(i) $\omega$-hyperon couplings
\begin{align}
\frac{g_{\omega \Lambda}}{g_{\omega N}}
= \frac{\sqrt{6} - 2z_v(1 - \alpha_v)}
       {\sqrt{6} + z_v(4\alpha_v - 1)},\\
\frac{g_{\omega \Sigma}}{g_{\omega N}}
= \frac{\sqrt{6} + 2z_v(1 - \alpha_v)}
       {\sqrt{6} + z_v(4\alpha_v - 1)},\\
\frac{g_{\omega \Xi}}{g_{\omega N}}
= \frac{\sqrt{6} - z_v(1 + 2\alpha_v)}
       {\sqrt{6} + z_v(4\alpha_v - 1)}.
\end{align}
(ii) $\phi$-hyperon couplings
\begin{align}
\frac{g_{\phi \Lambda}}{g_{\omega N}}
= \frac{-\sqrt{3} - 2\sqrt{2}z_v(1 - \alpha_v)}
       {\sqrt{6} + z_v(4\alpha_v - 1)},\\
\frac{g_{\phi \Sigma}}{g_{\omega N}}
= \frac{-\sqrt{3} + 2\sqrt{2}z_v(1 - \alpha_v)}
       {\sqrt{6} + z_v(4\alpha_v - 1)},\\
\frac{g_{\phi \Xi}}{g_{\omega N}}
= \frac{-\sqrt{3} - \sqrt{2}z_v(1 + 2\alpha_v)}
       {\sqrt{6} + z_v(4\alpha_v - 1)},\\
\frac{g_{\phi N}}{g_{\omega N}}
= \frac{-\sqrt{3} + \sqrt{2}z_v(4\alpha_v - 1)}
       {\sqrt{6} + z_v(4\alpha_v - 1)}.
\end{align}
(iii) $\rho$-hyperon couplings
\begin{align}
\frac{g_{\rho \Lambda}}{g_{\rho N}}
= 0, \\
\frac{g_{\rho \Sigma}}{g_{\rho N}}
= 2\alpha_v, \\
\frac{g_{\rho \Xi}}{g_{\rho N}}
= - (1 - 2\alpha_v).
\end{align}

When $\phi$ interaction is considered in SU(3) symmetry we must reparametrize $g_{\omega N}$ so that the nuclear matter properties remain unchanged \cite{Weissenborn:2011kb, 2014PhRvC..89b5805L}:

\begin{equation}
    g_{\omega N} \omega_0 \rightarrow\tilde{g}_{\omega N} \omega_0 + g_{\phi N} \phi_0.
\end{equation}
Then we obtain the modified value of $g_{\omega N}$ as
\begin{equation}
    \tilde{g}_{\omega N} = g_{\omega N} m_\phi \sqrt{\frac{1}{m_{\phi}^{2}+\big(\frac{-\sqrt{3} + \sqrt{2}z_v(4\alpha_v - 1)}{\sqrt{6} + z_v(4\alpha_v - 1)}\big)^2 m_{\omega}^2}}.
\end{equation}

\subsection{Bayesian analysis}

In Bayesian parameter estimation, the probability distribution of a set of
model parameters $\boldsymbol{\theta}$, referred to as the posterior probability density
function, conditioned on the data $D$, is obtained using Bayes'
theorem,
\begin{equation}
P(\boldsymbol{\theta}\mid D)
=
\frac{\mathcal{L}(D\mid\boldsymbol{\theta})\,\pi(\boldsymbol{\theta})}{Z}.
\end{equation}
Here, $\pi(\boldsymbol{\theta})$ denotes the prior probability distribution of the parameter
set $\boldsymbol{\theta}$, while the likelihood function $\mathcal{L}(D\mid\boldsymbol{\theta})$ incorporates the
information provided by experimental or observational data to update the
prior. The denominator $Z=P(D)$, known as the evidence, serves as a
normalization constant ensuring that the posterior distribution
$P(\boldsymbol{\theta}\mid D)$ is properly normalized.

In the present work, the parameter vector \(X\) consists of density-dependent couplings from the DDRMF framework.  To sample the posterior distribution, we employ the nested-sampling Monte
Carlo algorithm \texttt{MLFriends}~\cite{Buchner_2014,Buchner_2019} implemented in the \texttt{UltraNest} package~\cite{buchner2021ultranestrobustgeneral}. UltraNest is well-suited for complex inference problems, including multimodal posterior
distributions, nonlinear correlations among parameters, and posteriors
with heavy or light tails. Sampling is performed using the slice
sampler~\cite{10.1093/mnras/sty3090} available within UltraNest, which is
particularly efficient in high-dimensional parameter spaces and ensures
robust convergence during the sampling process. The number of slice-sampling steps is fixed through a series of nested-sampling runs with progressively increasing step counts, and the procedure is halted once the logarithm of the Bayesian evidence, $\log Z$ (marginal likelihood), shows convergence.

\begingroup

In addition to posterior sampling, nested sampling provides the Bayesian evidence, or marginal likelihood,
\begin{equation}
Z_M =
\int d\boldsymbol{\theta},
\mathcal{L}\left(D \mid \boldsymbol{\theta}, M\right),
\pi\left(\boldsymbol{\theta} \mid M\right),
\end{equation}
for a model $M$, where $\mathcal{L}$ denotes the likelihood, $\pi$ denotes the prior distribution, and $\boldsymbol{\theta}$ denotes the set of model parameters. The evidence quantifies the probability of the data under a given model after marginalizing over the full prior volume. It therefore incorporates both the quality of fit and the effective complexity of the model.

The relative support for two models, $M_i$ and $M_j$, is quantified through the Bayes factor,
\begin{equation}
B_{ij} =
\frac{Z_i}{Z_j},
\qquad
\Delta \ln Z_{ij} =
\ln Z_i - \ln Z_j =
\ln B_{ij}.
\end{equation}
A positive value of $\Delta \ln Z_{ij}$ indicates preference for model $M_i$ over model $M_j$. Following the Kass–Raftery interpretation, Bayes factors in the ranges $1<B_{ij}<3$, $3\leq B_{ij}<20$, $20\leq B_{ij}<150$, and $B_{ij}\geq150$ correspond, respectively, to evidence that is not worth more than a bare mention, positive, strong, and very strong in favor of model $M_i$ \cite{KassRaftery1995}. Equivalently, these ranges correspond approximately to $0<\Delta\ln Z<1.1$, $1.1\leq\Delta\ln Z<3.0$, $3.0\leq\Delta\ln Z<5.0$, and $\Delta\ln Z\geq5.0$. These categories are descriptive guidelines rather than strict decision thresholds, since the evidence can depend on the adopted prior ranges and modeling assumptions \cite{Trotta2008}.

\endgroup

\subsubsection{Parameters and priors}
The coupling constants of the RMF model are treated as free parameters, chosen such that the explored parameter space remains sufficiently broad. For the hyperonic sector in both the SU(6) and SU(3) scenarios, the relevant coupling ratios are also treated as free parameters and varied within the ranges listed in Table~\ref{tab:prior}. For the SU(6) case, we adopt the ranges of the scalar hyperon-coupling ratios suggested in Ref.~\cite{Huang:2024rvj}, whereas for the SU(3) case, where the corresponding couplings are less constrained, we allow them to vary over the full interval from 0 to 1. We further require that the corresponding hyperon optical potentials in symmetric nuclear matter remain within empirically motivated intervals, namely $-30 < U_{\Lambda} < -25$ MeV for the $\Lambda$ hyperon potential, $10 < U_{\Sigma} < 40$ MeV for the $\Sigma$ hyperon potential, and $-25 < U_{\Xi} < -10$ MeV for the $\Xi$ hyperon potential \cite{Huang:2024rvj}. In the SU(3) flavor symmetry scenario, in addition to the above parameters, the symmetry parameters $\alpha_v$ and the weight factor $z_v$, which represent the ratio of octet to singlet coupling, are also treated as free parameters. The parameter $\alpha_v = F/(F+D)$ quantifies the relative weight of antisymmetric to total couplings and is therefore intrinsically bounded within $0 \leq \alpha_v \leq 1$, corresponding to the limits of pure D- and pure F-type interactions. In addition, requiring the $\omega$-mediated interaction to remain repulsive for all baryons imposes non-negative vector couplings (e.g., $g_{\omega B}/g_{\omega N} \geq 0$), which constrains the mixing parameter to $0 \leq z_v \leq 2/\sqrt{6}\,(\approx 0.816)$. All these ranges are taken from Refs.~\cite{athira2025kaon,1984PrPNP..12..171D, Huang:2024rvj, Weissenborn_2012}. All model parameters are assigned uniform priors, as summarized in Table~\ref{tab:prior}.

\begin{table}[]
\caption{Prior ($P$) configuration adopted for the parameters of the DDRH model in this study. The terms ``Minimum'' and ``Maximum'' denote the lower and upper bounds of the corresponding prior distribution, respectively. The nucleonic-sector priors are common to all cases. The hyperonic-sector priors are listed separately for the SU(6) and SU(3) symmetry assumptions. The meson masses are fixed to $m_\sigma = 550\,\mathrm{MeV}$, $m_\omega = 783\,\mathrm{MeV}$, and $m_\rho = 763\,\mathrm{MeV}$ and are taken to be common across all cases.}
\label{tab:prior}
\begin{tabular}{@{}llll@{}}
\toprule
\toprule
Parameters & \multicolumn{1}{c}{Prior} & Minimum & Maximum \\
\midrule

$g_{\sigma N}$ & Uniform & \multicolumn{1}{c}{8.5} & \multicolumn{1}{c}{12.0} \\
$g_{\omega N}$ & Uniform & \multicolumn{1}{c}{9.5} & \multicolumn{1}{c}{14.0} \\
$g_{\rho N}$   & Uniform & \multicolumn{1}{c}{2.5} & \multicolumn{1}{c}{8.0} \\
$a_{\sigma}$   & Uniform & \multicolumn{1}{c}{0.0} & \multicolumn{1}{c}{0.20} \\
$a_{\omega}$   & Uniform & \multicolumn{1}{c}{0.0} & \multicolumn{1}{c}{0.20} \\
$a_{\rho}$     & Uniform & \multicolumn{1}{c}{0.0} & \multicolumn{1}{c}{1.0} \\

\addlinespace[0.6em]
\multicolumn{4}{c}{SU(6)} \\
\addlinespace[0.2em]

$r_{\sigma\Lambda}$ & Uniform & \multicolumn{1}{c}{0.58} & \multicolumn{1}{c}{0.63} \\
$r_{\sigma\Sigma}$  & Uniform & \multicolumn{1}{c}{0.42} & \multicolumn{1}{c}{0.54} \\
$r_{\sigma\Xi}$     & Uniform & \multicolumn{1}{c}{0.28} & \multicolumn{1}{c}{0.35} \\

\addlinespace[0.6em]
\multicolumn{4}{c}{SU(3)} \\
\addlinespace[0.2em]

$r_{\sigma\Lambda}$ & Uniform & \multicolumn{1}{c}{0.0} & \multicolumn{1}{c}{1.0} \\
$r_{\sigma\Sigma}$  & Uniform & \multicolumn{1}{c}{0.0} & \multicolumn{1}{c}{1.0} \\
$r_{\sigma\Xi}$     & Uniform & \multicolumn{1}{c}{0.0} & \multicolumn{1}{c}{1.0} \\
$\alpha_v$          & Uniform & \multicolumn{1}{c}{0.0} & \multicolumn{1}{c}{1.0} \\
$z_v$               & Uniform & \multicolumn{1}{c}{0.0} & \multicolumn{1}{c}{0.816} \\

\bottomrule
\end{tabular}
\end{table}

\begin{table*}
\caption{\label{tab:constraints}
Summary of the empirical, theoretical, and observational constraints used in this work.}
\begin{ruledtabular}
\begin{tabular}{lcccc}
\multicolumn{5}{c}{\textbf{(i) Nuclear matter properties}} \\[2pt]
Constraint & Density $n$ (fm$^{-3}$) & Observable & Value & Ref. \\[3pt]

\textit{Symmetric NM} \\
$n_0$    &     & Saturation density (fm$^{-3}$) & $0.153 \pm 0.005$ &  \\
$E_0$       & $n_0$     & Energy per nucleon (MeV)       & $-16.1 \pm 0.2$   &  \\
GMR                   & $n_0$            & $K_{0}$ (MeV)          & $230 \pm 30$   & \cite{PhysRevLett.95.122501}  \\[4pt]

\textit{Symmetry energy and symmetry pressure} \\
$\alpha_D$            & 0.05            & $J(n)$ (MeV)                 & $15.9 \pm 1.0$ & \cite{PhysRevC.92.031301} \\

IAS                   & $0.106\pm0.006$ & $J(n)$ (MeV)                 & $25.5 \pm 1.1$ & \cite{DANIELEWICZ2017147} \\[4pt]

\textit{Heavy-ion collisions (isospin-sensitive)} \\
HIC (Iso-diff)        & $0.035\pm0.011$ & $J(n)$ (MeV)                 & $10.3 \pm 1.0$ & \cite{PhysRevLett.102.122701} \\
HIC (n/p ratio)       & $0.069\pm0.008$ & $J(n)$ (MeV)                 & $16.8 \pm 1.2$ & \cite{MORFOUACE2019135045} \\
HIC ($\pi$ ratio)     & $0.232\pm0.032$ & $J(n)$ (MeV)                 & $52 \pm 13$    & \cite{PhysRevLett.126.162701} \\
HIC (n/p flow)        & $0.232$         & $P_{\rm sym}$ (MeV/fm$^3$)   & $10.9 \pm 8.7$ & \cite{RUSSOTTO2011471} \\[10pt]

\multicolumn{5}{c}{\textbf{(ii) $\chi$EFT constraints at low density}} \\[2pt]
$\chi$EFT PNM band & $n = 0.04$--$0.20$ fm$^{-3}$ & $E_{\rm SNM}(n)$, $E_{\rm PNM}(n)$ & 10\% expansion of  Ref.~\cite{PhysRevC.93.054314} & \cite{PhysRevC.93.054314} \\[10pt]

\multicolumn{5}{c}{\textbf{(iii) Astrophysical constraints}} \\[2pt]
Constraint & $M (M_\odot)$ & $R$ (km) & $\Lambda$ & Ref. \\[3pt]

\textit{Gravitational-wave observations} \\
GW170817 (LIGO/Virgo) & 1.4 &  & $190^{+390}_{-120}$ & \cite{PhysRevLett.119.161101} \\[4pt]

\textit{NICER pulsars} \\
PSR J0030+0451 & $1.34^{+0.15}_{-0.16}$ & $12.71^{+1.14}_{-1.19}$ & & \cite{Riley_2019} \\
 PSR J0030+0451 & $1.44^{+0.15}_{-0.14}$ & $13.02^{+1.24}_{-1.06}$ & & \cite{Miller_2019} \\
 PSR J0740+6620 & $2.07^{+0.07}_{-0.07}$ & $12.39^{+1.30}_{-0.98}$ & & \cite{Riley_2021} \\
 PSR J0740+6620 & $2.08^{+0.07}_{-0.07}$ & $13.70^{+2.6}_{-1.5}$    & & \cite{2021ApJ...918L..28M} \\
PSR J0437--4715 & $1.418^{+0.037}_{-0.037}$ & $11.36^{+0.95}_{-0.63} $  & & \cite{Choudhury_2024} \\

\end{tabular}
\end{ruledtabular}
\end{table*}

\subsubsection{Constraints}
Since the mid-1990s, considerable effort has been devoted to constraining the EOS of dense matter. Following Refs.~\cite {Huth_2022, Tsang_2024}, we employ a broad set of constraints spanning a wide range of densities, combining empirical nuclear matter properties, experimental data from heavy-ion collisions, microscopic calculations based on chiral effective field theory ($\chi$EFT), and astrophysical observations of NSs.

At saturation density, we adopt standard empirical constraints on symmetric nuclear matter, including the incompressibility $K_0 = 230 \pm 30$ MeV inferred from the Giant Monopole Resonance  \cite{PhysRevLett.95.122501}.  The density dependence of the symmetry energy, $J(n)$, is constrained by a range of complementary observables probing different density regimes. At low densities ($n \approx 0.05~\mathrm{fm}^{-3}$), the electric dipole polarizability of $^{208}$Pb provides symmetry energy constraints \cite{PhysRevC.92.031301}. In the intermediate region ($n \approx 0.10$--$0.12~\mathrm{fm}^{-3}$), nuclear mass systematics and isobaric analog states offer additional restrictions \cite{DANIELEWICZ2017147, PhysRevLett.111.232502, PhysRevC.85.024304}. Sub-saturation constraints are further informed by heavy-ion collision observables, such as isospin diffusion and neutron/proton spectral ratios \cite{PhysRevLett.102.122701, MORFOUACE2019135045}, while at densities near and above saturation, charged-pion ratios and neutron/proton elliptic-flow differences constrain the symmetry pressure \cite{PhysRevLett.126.162701, RUSSOTTO2011471}. 
In addition, we impose a super-Gaussian likelihood on the symmetry energy at saturation and its slope parameter, restricting $J(n_0)$ to the range $25$--$40$ MeV and $L$ to $30$--$120$ MeV, thereby ensuring physically reasonable behavior of the isovector sector \cite{parmar_2026}. A similar treatment is adopted for the saturation density $n_0$ and the energy per nucleon $E_0$, which are constrained around their empirical values; this also aids in achieving faster convergence of the Bayesian sampling.
At subsaturation densities, we also incorporate $\chi$EFT calculations for symmetric and pure neutron matter \cite{PhysRevC.93.054314}, with uncertainties conservatively enlarged by $\approx 10\%$ to account for theoretical systematics.

At higher densities, astrophysical observations provide essential constraints. We include tidal deformability measurements from the binary NS merger GW170817. Additional constraints are imposed using mass-radius measurements from PSR J0030+0451 \cite{Riley_2019,Miller_2019}, PSR J0740+6620 \cite{2021ApJ...918L..28M, Riley_2021} (with radio timing mass $M = 2.08 \pm 0.07\,M_{\odot}$ \cite{Fonseca_2021}), and PSR J0437--4715 \cite{Choudhury_2024}. These constraints are implemented using kernel density estimates of the posterior distributions from NICER measurements, which are incorporated as likelihood functions in the mass--radius plane within our Bayesian framework.
A summary of all empirical, experimental, and observational inputs employed in this work is provided in Table~\ref{tab:constraints}.

\begin{figure}
    \centering
    \includegraphics[width=.9\linewidth]{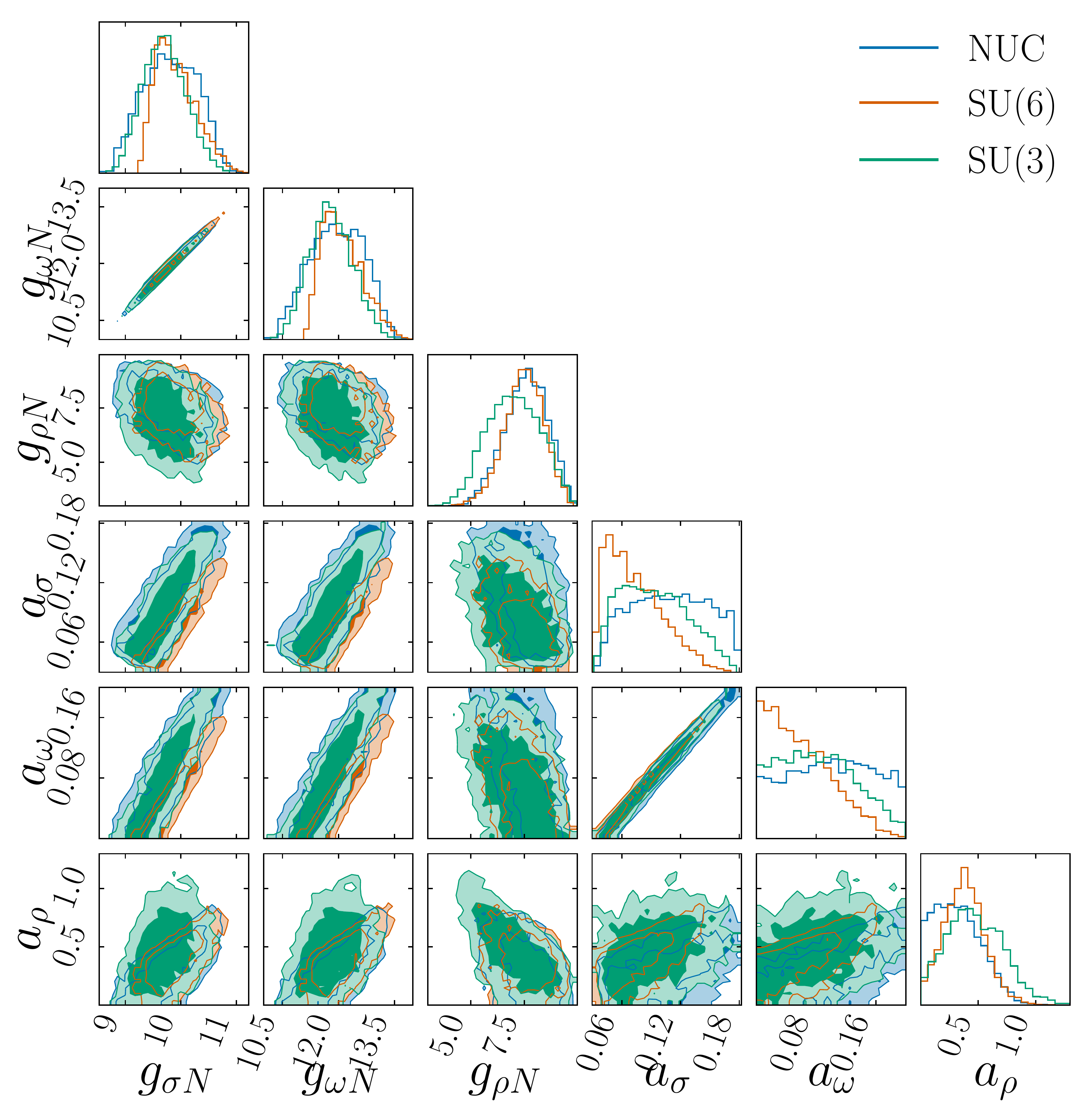}
    \caption{Marginalized posterior distributions of the DDRMF coupling constants $g_{\sigma N}$, $g_{\omega N}$, $g_{\rho N}$ and the density--dependence parameters $a_{\sigma}$, $a_{\omega}$, and $a_{\rho}$ obtained from the Bayesian analysis for the nucleonic, SU(6), and SU(3) symmetry schemes. The diagonal panels show the one-dimensional marginalized posteriors, while the off-diagonal panels display the two-dimensional joint posteriors. The contours represent the \(1\sigma\) and \(2\sigma\) credible regions.}
    \label{fig:commom_parameters}
\end{figure}

\section{Results}
\label{Results}

\subsection{Posterior distributions of model parameters}
The posterior distributions of the DDRMF model parameters $g_{\sigma}$, $g_{\omega}$, $g_{\rho}$, $a_{\sigma}$, $a_{\omega}$, and $a_{\rho}$ 
for the nucleonic, SU(6), and SU(3) symmetry schemes are presented in Fig.~\ref{fig:commom_parameters}. The diagonal panels show the one-dimensional marginalized posterior distributions of the 
individual parameters. The off-diagonal panels illustrate the two-dimensional joint posterior distributions, highlighting the correlations among the model parameters through the elliptical confidence contours. In particular, the coupling constants $g_{\sigma}$ and $g_{\omega}$ exhibit a strong correlation, reflecting the tight constraints imposed by the nuclear saturation properties, while moderate correlations are also observed between the coupling constants and their corresponding density dependence parameters $a_{\sigma}$ and $a_{\omega}$. From the plot, the marginalized one-dimensional posterior distributions of the model parameters are largely similar between the nucleonic and SU(3) cases, indicating that the inclusion of hyperons within the SU(3) framework does not substantially modify the preferred parameter space. In contrast, the SU(6) case exhibits noticeable deviations in the distributions for most parameters, reflecting a systematic shift in their inferred values. An exception to this trend is the $g_\rho$ coupling, whose distribution remains nearly unchanged across all three scenarios, signifying the dominant role of scalar $\sigma$ and isoscalar vector $\omega$ couplings in the hyperonic sector within the RMF framework.

\begin{figure}
    \centering
    \includegraphics[width=.9\linewidth]{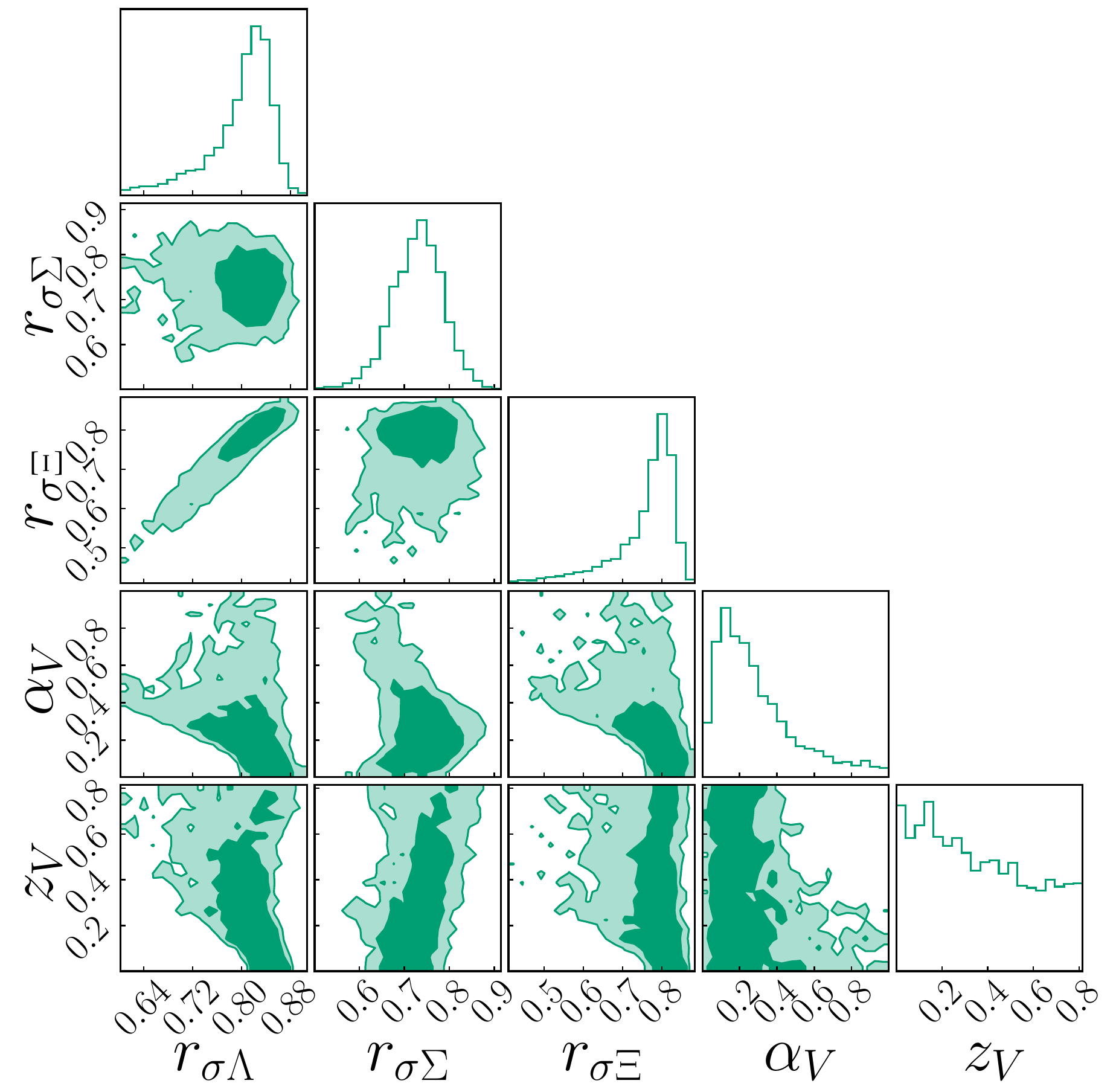}
    \caption{Marginalized posterior distributions of the hyperon coupling parameters obtained from the Bayesian analysis. Shown are the posterior distributions of the hyperon coupling ratios $r_{\sigma\Lambda}$, $r_{\sigma\Sigma}$, and $r_{\sigma\Xi}$, together with the SU(3) vector-coupling parameters $\alpha_v$ and $z_v$ for the SU(3) symmetry scheme.}
\label{fig:SU6_SU3}
\end{figure}

Figure~\ref{fig:SU6_SU3} provide a clear visualization of the posterior distributions and the correlations among the hyperon coupling parameters within the SU(3) symmetry framework.  In the SU(3) scenario, the scalar coupling ratios $r_{\sigma \Lambda}$, $r_{\sigma \Sigma}$, and $r_{\sigma \Xi}$ display relatively broad posterior distributions as compared to the SU(6), indicating that these parameters are only weakly constrained by the imposed physical conditions. This spread reflects a significant degree of uncertainty and flexibility in the scalar sector, allowing multiple combinations of couplings to satisfy the same observational or theoretical constraints. In contrast, the vector-sector parameters, particularly $\alpha_v$, appear comparatively better constrained, with narrower distributions suggesting a stronger sensitivity to the underlying constraints. Notably, $\alpha_v$ is fixed to $1$ in the SU(6) limit, whereas in the SU(3) case it exhibits a substantial deviation from this value, reflecting the additional freedom permitted by the relaxed symmetry. The two-dimensional projections further reveal nontrivial correlations among the scalar couplings, as well as between scalar and vector parameters. In particular, $r_{\sigma\Lambda}$ and $r_{\sigma\Xi}$ exhibit a strong positive correlation.

%Such correlations point to an inherent degeneracy in the model parameter space, wherein variations in one parameter can be compensated by adjustments in others, thereby preserving the overall physical consistency of the EOS.

On the other hand, the SU(6) symmetry case exhibits markedly different behavior. The posterior distributions of the hyperon coupling ratios are essentially flat, reflecting the adoption of flat priors and the fact that current observations do not provide significant constraints on these parameters, unlike in the SU(3) case. For this reason, these distributions are not shown. The corresponding median values, along with the 95\% ($2\sigma$) and 68\% ($1\sigma$) credible intervals, are reported as superscript/subscript bounds, with the latter given in parentheses, and are presented in Table~\ref{tab:hyperon_ci}. The values of $r_{\sigma\Lambda}$, $r_{\sigma\Sigma}$, and $r_{\sigma\Xi}$ are generally higher in the SU(3) case than in the SU(6) case. While SU(6) corresponds to $\alpha_v = 1$, the SU(3) posterior favors comparatively lower values of $\alpha_v$, with $z_v$ remaining allowed over a broad range of values with no correlation between them.

%Furthermore, the correlations observed in the SU(3) case are largely suppressed under SU(6), suggesting that the stronger symmetry assumptions effectively reduce parameter degeneracies and lead to a more rigid and predictive framework.

%Overall, the comparison between the two symmetry schemes highlights a clear trade-off between flexibility and predictability. The SU(3) framework, with its weaker symmetry constraints, allows for a broader exploration of the parameter space and accommodates a wider range of coupling combinations, albeit at the cost of increased degeneracy and uncertainty. In contrast, SU(6) symmetry imposes stricter relations among the couplings, resulting in more precise but less flexible parameter estimates.

\begin{table}
\caption{Median values with 95\% credible intervals (superscript/subscript) and 68\% credible intervals (in parentheses) for hyperon coupling ratios and vector interaction parameters.}
\centering
\renewcommand{\arraystretch}{1.5}

\begin{tabular}{lcc}
\hline
Parameter & \textbf{SU(6)} & \textbf{SU(3)} \\
\hline

$r_{\sigma\Lambda}$ & 
$0.602^{+0.026}_{-0.021}\,(^{+0.018}_{-0.015})$ & 
$0.811^{+0.055}_{-0.259}\,(^{+0.032}_{-0.075})$ \\

$r_{\sigma\Sigma}$ & 
$0.477^{+0.059}_{-0.054}\,(^{+0.041}_{-0.039})$ & 
$0.729^{+0.107}_{-0.216}\,(^{+0.053}_{-0.063})$ \\

$r_{\sigma\Xi}$ & 
$0.308^{+0.040}_{-0.027}\,(^{+0.027}_{-0.020})$ & 
$0.782^{+0.064}_{-0.498}\,(^{+0.038}_{-0.116})$ \\

$\alpha_V$ & 
--- & 
$0.242^{+0.677}_{-0.206}\,(^{+0.304}_{-0.139})$ \\

$z_V$ & 
--- & 
$0.343^{+0.450}_{-0.327}\,(^{+0.318}_{-0.241})$ \\

\hline
\end{tabular}
\label{tab:hyperon_ci}
\end{table}

\begin{table*}
\centering
\renewcommand{\arraystretch}{1.}
\setlength{\tabcolsep}{10pt}
\caption{Median values and associated credible intervals for the inferred nuclear matter and neutron star properties. For each quantity, we report the median together with the 68\% credible interval as superscript/subscript uncertainties, while the values in parentheses correspond to the 95\% credible interval.}
\begin{tabular}{lccc}
\hline\hline
Parameter & NUC & SU(6) & SU(3) \\
\hline

\multicolumn{4}{c}{\textbf{Nuclear matter properties}} \\[4pt]

$K_0$ (MeV) &
$257.321^{+28.201}_{-22.541}\,(^{+52.648}_{-39.618})$ &
$304.059^{+17.944}_{-14.260}\,(^{+31.240}_{-24.978})$ &
$281.127^{+24.558}_{-25.195}\,(^{+40.640}_{-45.108})$ \\[4pt]

$Q_0$ (MeV)&
$-45.957^{+193.459}_{-150.415}\,(^{+362.303}_{-246.059})$ &
$268.934^{+92.007}_{-106.078}\,(^{+173.085}_{-260.856})$ &
$77.224^{+165.398}_{-157.896}\,(^{+286.410}_{-276.867})$ \\[4pt]

$Z_0$(MeV) &
$740.578^{+616.862}_{-869.918}\,(^{+877.216}_{-1843.797})$ &
$603.032^{+598.227}_{-1112.023}\,(^{+794.466}_{-2338.826})$ &
$608.198^{+598.514}_{-945.382}\,(^{+917.822}_{-2111.269})$ \\[4pt]

$J$ (MeV)&
$32.467^{+3.746}_{-3.463}\,(^{+6.663}_{-6.081})$ &
$32.173^{+3.376}_{-3.052}\,(^{+6.776}_{-5.801})$ &
$30.885^{+5.059}_{-3.932}\,(^{+8.479}_{-6.136})$ \\[4pt]

$L$ (MeV)&
$67.994^{+19.090}_{-16.599}\,(^{+34.970}_{-29.027})$ &
$64.197^{+14.440}_{-12.339}\,(^{+29.651}_{-23.140})$ &
$57.017^{+21.256}_{-16.000}\,(^{+41.903}_{-25.327})$ \\[4pt]

$K_{\rm sym}$ (MeV)&
$-77.868^{+49.232}_{-38.416}\,(^{+89.473}_{-73.174})$ &
$-75.284^{+39.424}_{-33.384}\,(^{+88.368}_{-62.900})$ &
$-70.245^{+43.659}_{-38.523}\,(^{+91.599}_{-75.607})$ \\[4pt]

$Q_{\rm sym}$ (MeV)&
$243.488^{+373.776}_{-236.554}\,(^{+749.328}_{-327.173})$ &
$374.999^{+322.271}_{-300.582}\,(^{+654.305}_{-445.769})$ &
$461.778^{+308.604}_{-357.638}\,(^{+601.502}_{-535.822})$ \\[4pt]

$Z_{\rm sym}$ (MeV)&
$-2342.364^{+1031.238}_{-2611.533}\,(^{+1422.346}_{-6606.191})$ &
$-3815.124^{+1374.019}_{-2642.035}\,(^{+1855.431}_{-5969.732})$ &
$-4135.587^{+2114.786}_{-3783.638}\,(^{+2836.890}_{-7474.778})$ \\[8pt]

\multicolumn{4}{c}{\textbf{Neutron star properties}} \\[4pt]

$M_{\max}$ $(M_\odot)$ &
$2.222^{+0.114}_{-0.096}\,(^{+0.197}_{-0.189})$ &
$2.044^{+0.042}_{-0.031}\,(^{+0.083}_{-0.041})$ &
$2.183^{+0.097}_{-0.103}\,(^{+0.188}_{-0.168})$ \\[4pt]

$R_{\max}$ (km) &
$11.340^{+0.335}_{-0.295}\,(^{+0.630}_{-0.570})$ &
$11.675^{+0.345}_{-0.280}\,(^{+0.730}_{-0.475})$ &
$11.460^{+0.290}_{-0.290}\,(^{+0.580}_{-0.565})$ \\[4pt]

$R_{1.4}$ (km) &
$12.712^{+0.322}_{-0.322}\,(^{+0.625}_{-0.672})$ &
$12.987^{+0.273}_{-0.250}\,(^{+0.532}_{-0.450})$ &
$12.587^{+0.324}_{-0.285}\,(^{+0.679}_{-0.581})$ \\[4pt]

$\Lambda_{1.4}$ &
$586.746^{+108.541}_{-90.754}\,(^{+207.870}_{-165.434})$ &
$703.809^{+82.894}_{-80.807}\,(^{+207.764}_{-137.761})$ &
$559.096^{+83.690}_{-75.077}\,(^{+190.649}_{-142.201})$ \\[4pt]

$n_c(M_{\max})$(fm$^{-3}$) &
$0.973^{+0.071}_{-0.074}\,(^{+0.142}_{-0.128})$ &
$0.933^{+0.030}_{-0.039}\,(^{+0.059}_{-0.075})$ &
$0.952^{+0.054}_{-0.058}\,(^{+0.116}_{-0.109})$ \\[4pt]

$P_c(M_{\max})$ (MeV fm$^{-3}$)&
$502.640^{+46.485}_{-53.006}\,(^{+80.918}_{-90.809})$ &
$344.787^{+39.462}_{-50.352}\,(^{+66.836}_{-102.653})$ &
$441.838^{+51.009}_{-65.327}\,(^{+90.210}_{-126.334})$ \\[4pt]

$c_s^2(n_c)$ &
$0.659^{+0.060}_{-0.086}\,(^{+0.089}_{-0.139})$ &
$0.493^{+0.035}_{-0.058}\,(^{+0.053}_{-0.126})$ &
$0.601^{+0.065}_{-0.086}\,(^{+0.092}_{-0.168})$ \\[4pt]

$f_{1.4}\ (\mathrm{kHz})$ &
$1.682^{+0.056}_{-0.055}\,(^{+0.119}_{-0.103})$ &
$1.621^{+0.039}_{-0.042}\,(^{+0.073}_{-0.083})$ &
$1.692^{+0.052}_{-0.051}\,(^{+0.105}_{-0.107})$ \\[4pt]

$p_{1,1.4}\ (\mathrm{kHz})$ &
$6.166^{+0.276}_{-0.267}\,(^{+0.550}_{-0.527})$ &
$6.252^{+0.258}_{-0.234}\,(^{+0.487}_{-0.431})$ &
$6.469^{+0.385}_{-0.384}\,(^{+0.646}_{-0.682})$ \\

\hline\hline
\end{tabular}

%\caption{Median values with 95\% credible intervals (superscript/subscript) and 68\% credible intervals (in parentheses) for nuclear matter and neutron star properties.}
\label{tab:combined_summary}
\end{table*}

\begin{figure}
    \centering
    \includegraphics[width=0.96\linewidth]{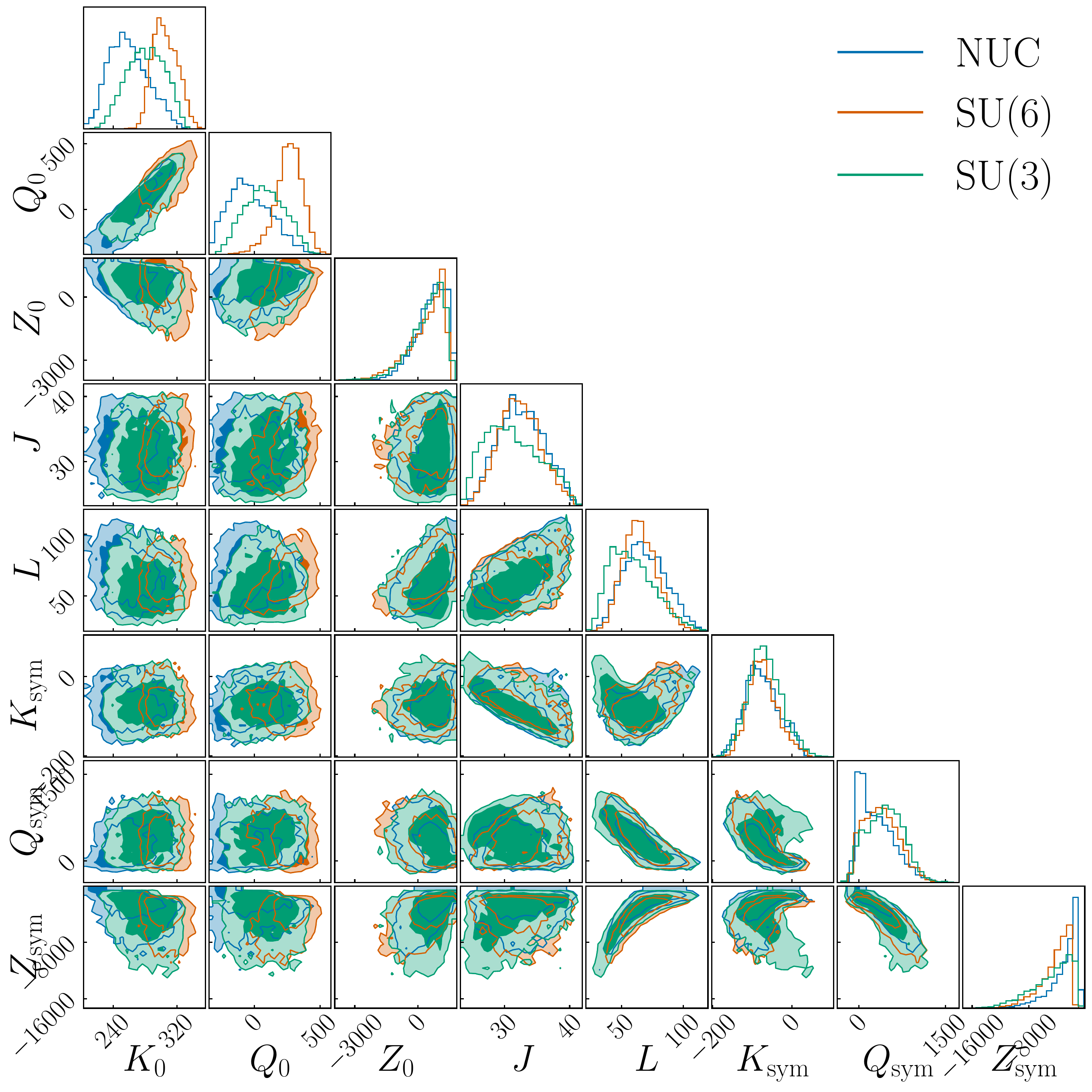}
    \caption{Posterior distributions of the nuclear matter parameters $K_0$, $Q_0$, $Z_0$, $J$, $L$, $K_{\rm sym}$, $Q_{\rm sym}$, and $Z_{\rm sym}$ for the nucleonic, SU(6), and SU(3) symmetry schemes. }
    \label{fig:nuc_prop}
\end{figure}

\begin{figure}
    \centering
    \includegraphics[width=1\linewidth]{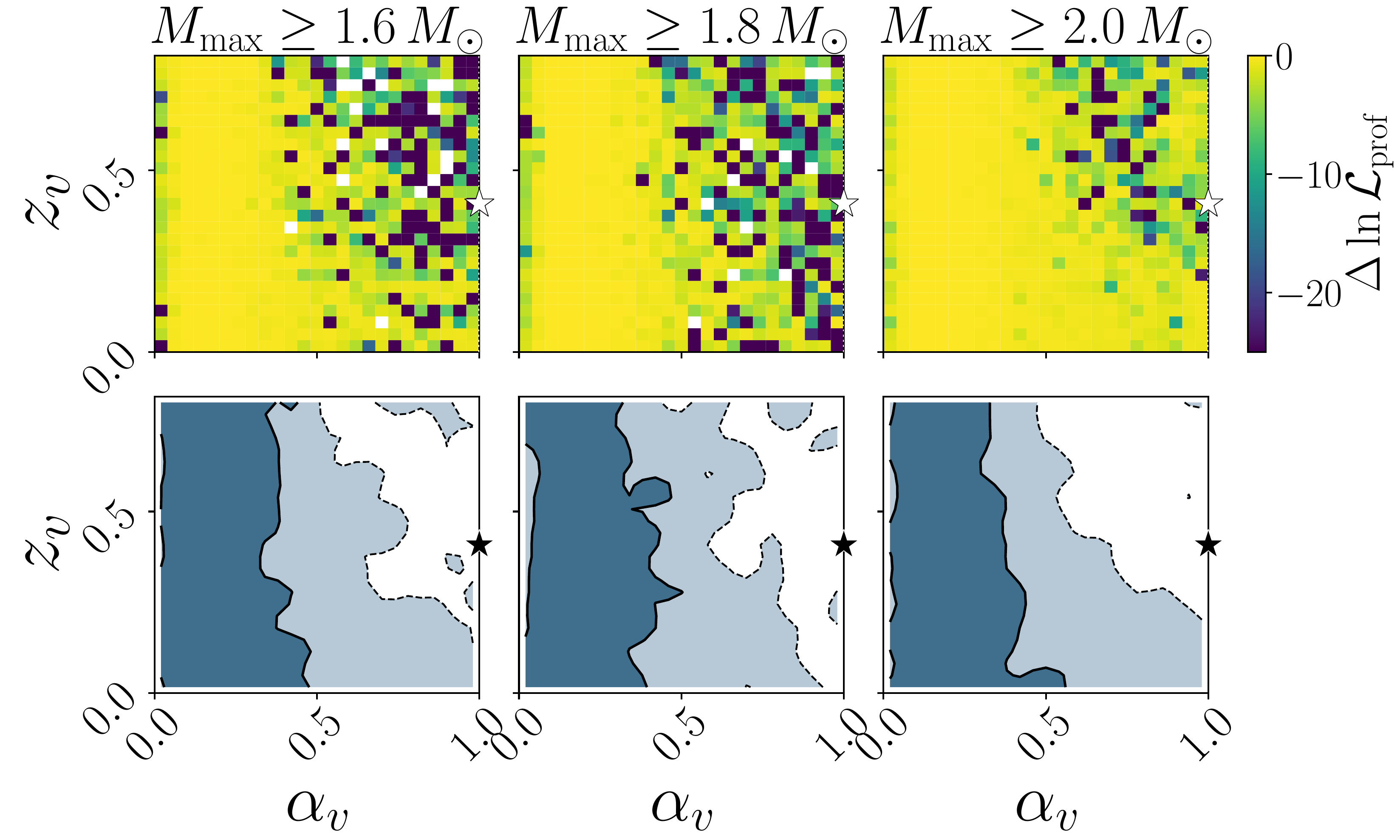}
   
    \caption{Sensitivity of the SU(3) vector-sector inference to the lower bound imposed on the maximum neutron-star mass. Upper panels show the relative profile likelihood in the $(\alpha_v,z_v)$ plane for the three adopted mass thresholds. Each upper panel is normalized independently to its own maximum likelihood. Lower panels show the corresponding marginalized posterior distributions after integrating over all remaining model parameters. Dark and light shaded regions enclose the 68\% and 95\% highest-posterior-density regions, respectively; solid and dashed curves indicate their boundaries. The star denotes the SU(6) reference point, $(\alpha_v,z_v)=(1,1/\sqrt{6})$, in the adopted parametrization.}
    \label{fig:masscut_sensitivity}
\end{figure}

\begin{figure*}
    \centering
    \includegraphics[width=0.8\linewidth]{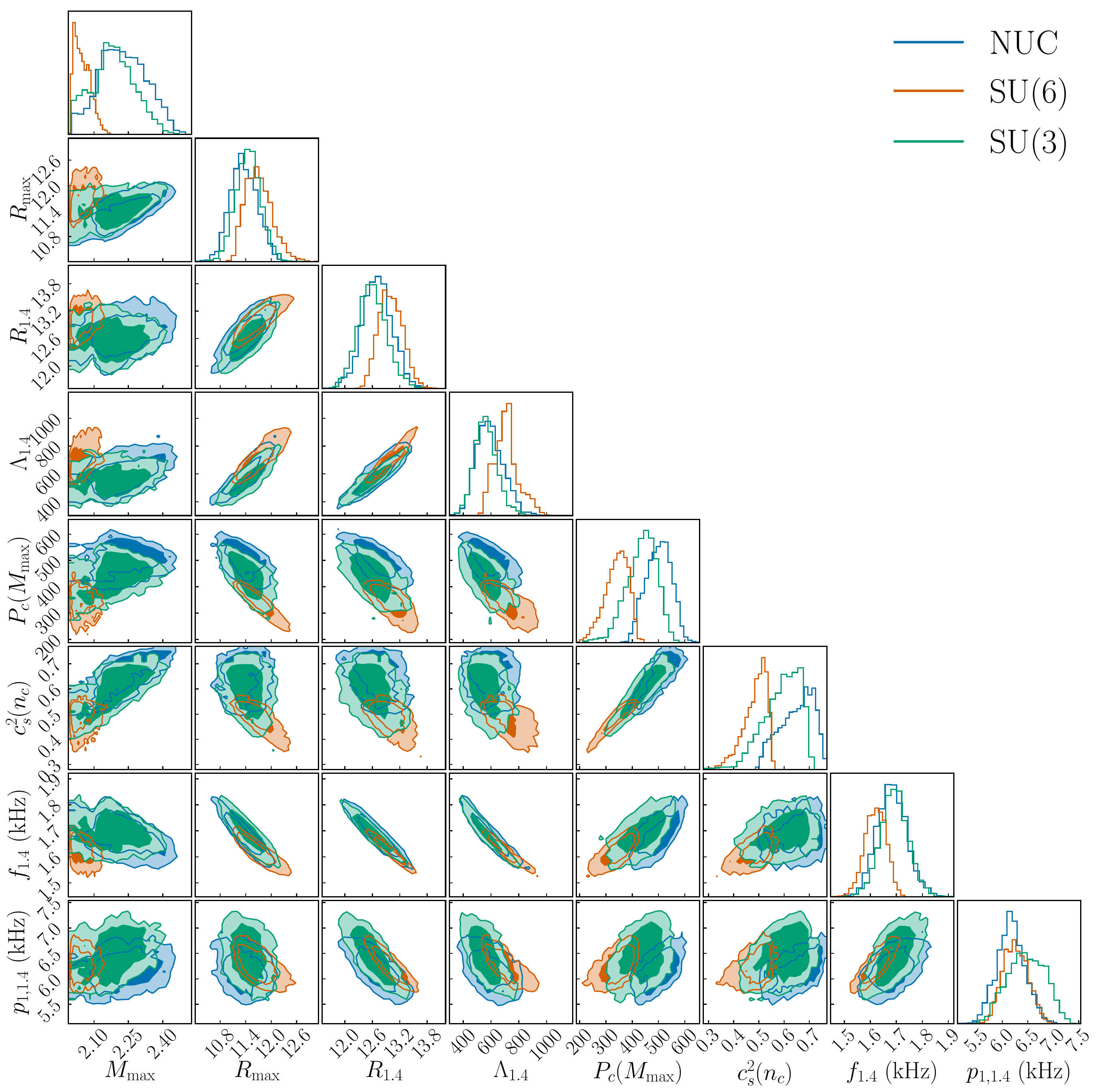}
    \caption{Posterior distributions of the NS observables: Maximum mass ($M_{\rm max}$), radius corresponding to maximum mass $R_{\rm max}$ in km, radius of the canonical star $R_{1.4}$ in km, tidal deformability of the canonical star $\Lambda_{1.4}$, central baryon density $n_c(M_{\rm max})$ in fm$^{-3}$, central pressure $P_c(M_{\rm max})$ in MeVfm$^{-3}$, sound speed $c_s(n_c)$,  $f$-mode and the $p$-mode frequencies of the canonical star in kHz are obtained from the Bayesian analysis for the nucleonic, SU(6), and SU(3) symmetry schemes.}
    \label{fig:ns_observables}
\end{figure*}

Figure~\ref{fig:nuc_prop} presents the posterior distributions and correlations among the key nuclear matter properties, including the incompressibility modulus $K_{0}$, the skewness parameter $Q_{0}$, and the symmetry-energy parameters $J$, $L$, $K_{\mathrm{sym}}$, $Q_{\mathrm{sym}}$, together with the higher-order coefficients $Z_{0}$ and $Z_{\mathrm{sym}}$. These quantities enter the expansion of the energy per particle of symmetric nuclear matter, $E(n)/A$, and of the symmetry energy, $S(n)$, around the saturation density $n_{0}$ as
\begin{equation}
\frac{E(n)}{A}
=
E_{0}
+\frac{K_{0}}{2!}\chi^{2}
+\frac{Q_{0}}{3!}\chi^{3}
+\frac{Z_{0}}{4!}\chi^{4}
+\cdots ,
\end{equation}
\begin{equation}
S(n)
=
J
+L\chi
+\frac{K_{\mathrm{sym}}}{2!}\chi^{2}
+\frac{Q_{\mathrm{sym}}}{3!}\chi^{3}
+\frac{Z_{\mathrm{sym}}}{4!}\chi^{4}
+\cdots ,
\end{equation}
with $\chi=(n-n_{0})/(3n_{0})$, where $S(n)$ denotes the symmetry energy as a function of baryon density. 
The expected correlations among the nuclear matter parameters are clearly visible in Fig.~\ref{fig:nuc_prop}. On the isoscalar side, the posterior distributions of $K_{0}$ and $Q_{0}$ differ noticeably when going from the nucleonic case to the SU(3) and SU(6) hyperonic scenarios. The SU(3) results remain comparatively close to the nucleonic ones, indicating that this framework largely preserves nucleonic-like behavior. In contrast, the SU(6) case exhibits a more pronounced shift, reflecting its tendency to generate softer equations of state, which in turn favors relatively larger values of the incompressibility and $Q_{0}$. The posterior distribution of $Z_{0}$, however, remains broadly similar in all three cases. By comparison, the isovector sector, namely $J$, $L$, $K_{\mathrm{sym}}$, $Q_{\mathrm{sym}}$, and $Z_{\mathrm{sym}}$, is almost independent of the adopted model, whether nucleonic, SU(3), or SU(6). This is in line with earlier Bayesian studies indicating that isovector nuclear matter properties remain largely unaffected by the specific hyperonic prescription \cite{Huang:2024rvj, PhysRevD.106.063024}.

\begingroup

To assess whether the inferred SU(3) vector-sector preference is mainly driven by the adopted maximum-mass requirement or by the full set of constraints entering the analysis, we repeated the SU(3) inference with otherwise identical priors and likelihood terms, imposing $M_{\rm max}\geq1.6\,M_\odot$, $1.8\,M_\odot$, and $2.0\,M_\odot$. Figure~\ref{fig:masscut_sensitivity} presents two complementary views of the resulting $(\alpha_v,z_v)$ constraints.

The upper panels show the sampled relative profile likelihood,
\begin{equation}
\Delta\ln\mathcal{L}_{\rm prof}(\alpha_v,z_v)
=
\ln\mathcal{L}_{\rm prof}(\alpha_v,z_v)
-
\ln\mathcal{L}_{\rm max},
\end{equation}
where $\ln\mathcal{L}_{\rm prof}(\alpha_v,z_v)$ is the largest likelihood found among the sampled models within a given $(\alpha_v,z_v)$ bin, and $\ln\mathcal{L}_{\rm max}$ is the largest likelihood in the corresponding analysis. We show this quantity to determine whether a region of the vector-sector parameter space contains at least one model that can fit the full dataset competitively after the remaining RMF parameters are allowed to vary.  The broader near-maximum regions for the stricter mass cuts indicate that the maximum-mass requirement primarily removes strongly soft equations of state, while leaving a residual degeneracy among several viable $(\alpha_v,z_v)$ combinations. Thus, the high-mass condition alone does not uniquely determine the SU(3) vector-sector parameters. The lower panels show   that \% and 95\% highest-posterior-density regions remain broadly stable as the lower bound on $M_{\rm max}$ is varied from $1.6\,M_\odot$ to $2.0\,M_\odot$, with no systematic shift of the preferred vector-sector region. Hence, within the present DDRMF framework, the inferred preference for SU(3) couplings differing from the restrictive SU(6) point is not produced solely by the precise choice of the maximum-mass cutoff. Instead, it is a conditional result of the combined nuclear, hypernuclear, gravitational-wave, and NICER constraints, together with the adopted priors and model parametrization. This test does not imply that the generic requirement of sufficient high-density stiffness is irrelevant; it shows that tightening the adopted lower bound on $M_{\rm max}$ over the range considered does not qualitatively alter the inferred vector-sector posterior.

\endgroup

The posterior distribution of  NS observables is presented in  Fig.~\ref{fig:ns_observables}. The well-known correlations among NS observables, such as those involving $R_{1.4}$ and $\Lambda_{1.4}$, are recovered here as well. As in the nuclear matter sector, the SU(3) case again shows distributions that remain very similar to those of the purely nucleonic scenario, whereas the SU(6) case exhibits more distinct behavior. In particular, the posterior distribution of $M_{\mathrm{max}}$ in the SU(3) framework remains close to that of the nucleonic case, while in SU(6) it is more narrowly concentrated around $\approx 2.04\, M_{\odot}$. This suggests that the additional flexibility available in the SU(3) case through the parameters $\alpha_v$ and $z_v$ allows the model to support comparatively larger NS masses while still preserving nucleonic-like behavior. A similar trend is seen for the stellar radius: SU(3) gives values of $R_{1.4}$ that are broadly consistent with the nucleonic case, whereas SU(6) tends to favor larger canonical radii.  The $f$-mode frequencies and damping times are computed in full general relativity using the publicly available \texttt{\url{}} code, which solves the quadrupolar \((l=2)\) fluid perturbations of a static NS on top of the TOV background by direct numerical integration of the perturbed Einstein--fluid equations. The interior perturbation equations are integrated from the center and the surface and matched at an intermediate point, after which the exterior Zerilli equation is solved, and a purely outgoing-wave condition is imposed to determine the complex eigenfrequency, whose real and imaginary parts give the mode frequency and damping time, respectively \cite{f_mode}. 

The same pattern extends to the oscillation properties computed in full general relativity. The $f$-mode frequencies are generally reduced in the SU(6) framework compared with both the nucleonic and SU(3) cases, while the $p_{1}$-mode frequencies remain almost unchanged across the different prescriptions. These results indicate that, when quantifying the impact of hyperons on NS bulk and oscillation properties, the underlying symmetry prescription plays a particularly important role. In this sense, not only the presence of hyperons, but also the specific symmetry structure used to determine their couplings becomes essential for characterizing the resulting EOS and its astrophysical predictions.
%A strong positive correlation is observed between the radii $R_{1.4}$ and $R_{\mathrm{max}}$, indicating that equations of state predicting larger radii for canonical $1.4\, M_{\odot}$ stars also tend to yield larger radii for the maximum-mass configuration. Similarly, the tidal deformability $\Lambda_{1.4}$ shows a clear positive correlation with $R_{1.4}$, reflecting the well-known sensitivity of tidal deformability to the stellar radius. The maximum mass $M_{\mathrm{max}}$ exhibits moderate correlations with central properties such as the central density $n_{c}(M_{\mathrm{max}})$ and central pressure $P_{c}(M_{\mathrm{max}})$, where larger maximum masses are generally associated with lower central densities and higher central pressures, indicating stiffer equations of state. An inverse correlation is also noticeable between $M_{\mathrm{max}}$ and $n_{c}(M_{\mathrm{max}})$, suggesting that more massive stars can be supported at comparatively lower central densities. The speed of sound at the central density, $c_{s}(n_{c})$, shows correlations with both $M_{\mathrm{max}}$ and central pressure, further highlighting its role in determining the stiffness of the EOS at high densities. Overall, the SU(6) case exhibits more compact and tightly clustered contours compared to SU(3), indicating reduced uncertainties and weaker degeneracies among the neutron star observables, while the SU(3) framework allows for a broader range of correlated variations.
For completeness, the median values of the nuclear matter and neutron star properties are summarized in Table \ref{tab:combined_summary}, where both the 95\% and 68\% credible intervals are reported for each quantity across the three cases: nucleonic matter, SU(6), and SU(3) frameworks.

\begin{figure}
    \centering
    \includegraphics[width=\linewidth]{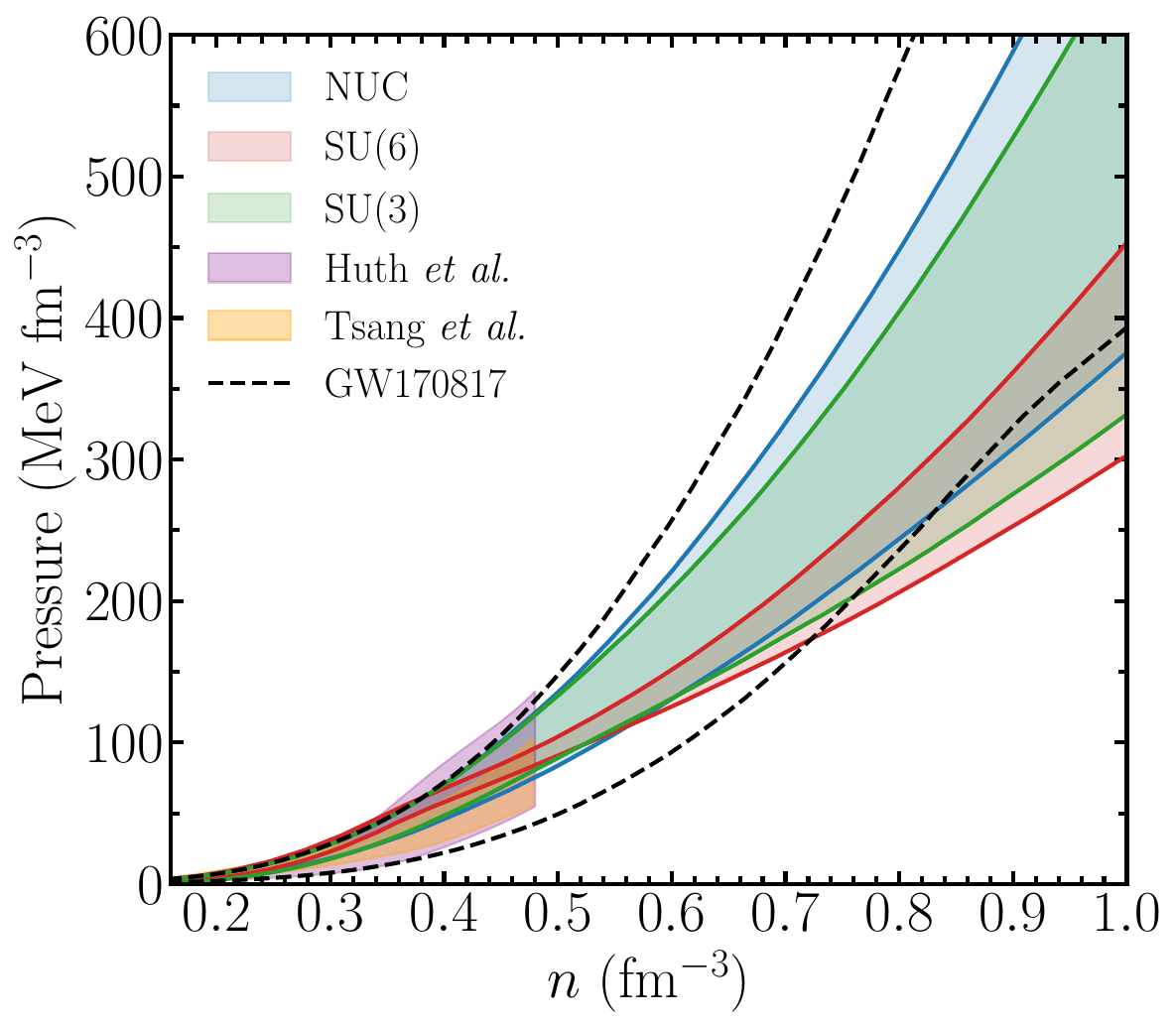}
    \caption{Pressure as a function of baryon density for the EOSs corresponding to the nucleonic, SU(6), and SU(3) symmetry schemes. For comparison, we also show the constraints reported by Huth \textit{et al.} \cite{Huth_2022}, the multiphysics constrained band from Tsang \textit{et al.} \cite{Tsang_2024}, and the pressure constraints inferred from the gravitational-wave event GW170817 \cite{PhysRevLett.119.161101}.}
    \label{fig:PvsRho}
\end{figure}

Having analyzed the distributions of the coupling parameters for all cases, we now proceed to examine the implications of the symmetry framework on the macroscopic properties of NSs, beginning with the EOS. Figure~\ref{fig:PvsRho} presents the constraints on the EOS derived from our Bayesian analysis of the EOSs for nucleonic, SU(6), and SU(3) symmetry schemes. The resulting pressure–density band mostly shows agreement with gravitational-wave observations from the GW170817 event \cite{PhysRevLett.119.161101}, as well as with representative nuclear-theory predictions, such as those reported by Huth \textit{et al.} \cite{Huth_2022} and Tsang \textit{et al.} \cite{Tsang_2024}, across the entire density range considered. Clearly, the posterior distribution of the SU(3) EOS closely resembles that of the purely nucleonic case. The nucleonic EOS remains the stiffest, while the inclusion of hyperons leads to a softening in both SU(3) and SU(6) scenarios \cite{2014PhRvC..89b5805L, Glendenning:1997wn, Haensel:2007yy, WEBER1989779, Rufa:1987uf}. Among these, the SU(6) case exhibits the softest behavior and fails to remain compatible with the constraints from the GW170817 event at higher densities. The relatively stiff behavior found in the SU(3) scenario is particularly interesting in view of current discussions of the hyperon puzzle. It is qualitatively in line with modern microscopic and \textit{ab initio} calculations \cite{PhysRevLett.114.092301, Logoteta_2019, Tong_2025}, which indicate that the inclusion of repulsive many-body interactions in the strange sector can significantly reduce the softening associated with hyperon onset to support NS above $2\, M_{\odot}$. From this perspective, the SU(3) case may be interpreted as reflecting a hyperonic EOS in which the balance between attraction and repulsion remains favorable to massive NS configurations.

\begin{figure*}
    \centering
    \includegraphics[width=1\linewidth]{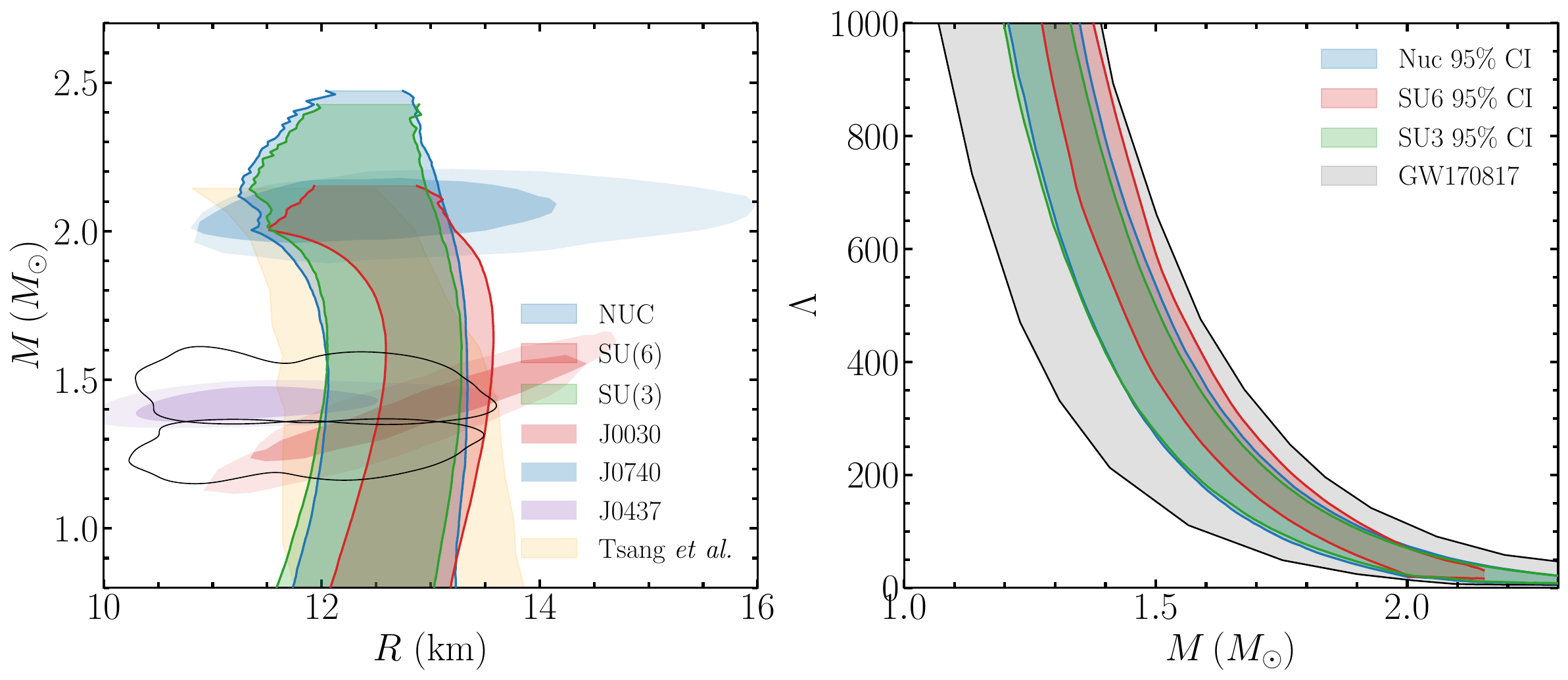}
    \caption{Right panel: Maximum mass as a function of radius for the EOSs corresponding to the nucleonic, SU(6), and SU(3) symmetry schemes. For comparison, we also show the multiphysics constrained band from Tsang \textit{et al.} \cite{Tsang_2024}, and the shaded elliptical regions depict the mass-radius (\textit{M-R}) constraints derived from NICER pulse-profile analyses of PSR J0030+0451 and PSR J0740+6620. The darker and lighter bands correspond to the 68\% and 95\% credibility intervals, respectively. An additional constraint at lower masses is provided by PSR J0437--4715. Left panel:  Lambda deformability as a function of maximum mass for the EOSs corresponding to the nucleonic, SU(6), and SU(3) symmetry schemes. The contours represent the \textit{M-R} posterior inferred from the GW170817 event \cite{PhysRevLett.119.161101}. The blue shaded band represents the nucleonic case, the red band corresponds to the SU(6) symmetry scheme, and the green band denotes the SU(3) symmetry case.}
    \label{fig:MR_lambda}
\end{figure*}

In Fig.~\ref{fig:MR_lambda}, the left panel displays the M-R relations obtained for all the considered cases. For comparison, constraints from pulse-profile modeling by the NICER mission are overlaid, including results for PSR J0030+0451, PSR J0740+6620, and PSR J0437$-$4715, where the darker and lighter shaded regions correspond to the 68\% and 95\% credible intervals, respectively. The predicted EOSs predominantly lie within the 68\% and 95\% credible regions reported by Tsang \textit{et al.} \cite{Tsang_2024} in the M-R plane, indicating consistency with theikernel r analysis.
The right panel of Fig.~\ref{fig:MR_lambda} shows the dimensionless tidal deformability ($\Lambda$) as a function of NS mass. The 95\% credible intervals corresponding to the three cases are presented, and the inferred values lie well within the bounds established by the gravitational-wave observation of GW170817~\cite{PhysRevLett.119.161101}, thereby supporting the overall compatibility of our results with multimessenger astrophysical constraints.
For both plots, the SU(6) case exhibits a trend consistent with results reported in the literature~\cite{PhysRevD.106.063024}.  As observed earlier, the nucleonic and SU(3) cases show similar behavior, with their corresponding curves closely overlapping in both the \textit{M-R} and \textit{$\Lambda$-M} planes. The recent NICER measurement of PSR~J0437$-$4715, which gives a mass of $M=1.418\pm0.037\, M_{\odot}$ and an equatorial radius of $R=11.36^{+0.95}_{-0.63}\,\mathrm{km}$, points toward a comparatively soft EOS around the canonical-mass regime \cite{Choudhury_2024}. In this context, our results suggest that such a compact $\approx 1.4\, M_{\odot}$ star is difficult to reconcile with the SU(6) symmetry prescription, whereas the SU(3) case remains compatible with both this NICER inference and the broader astrophysical constraints shown in Fig.~\ref{fig:MR_lambda}. This is particularly important because relatively small radii in the $1.4\, M_{\odot}$ mass range have long been regarded as a possible discriminator of hyperonic softening. Our analysis shows that such an interpretation depends crucially on the assumed symmetry scheme and that hyperons need not be disfavored.

\begin{figure*}
    \centering
    \includegraphics[width=\linewidth]{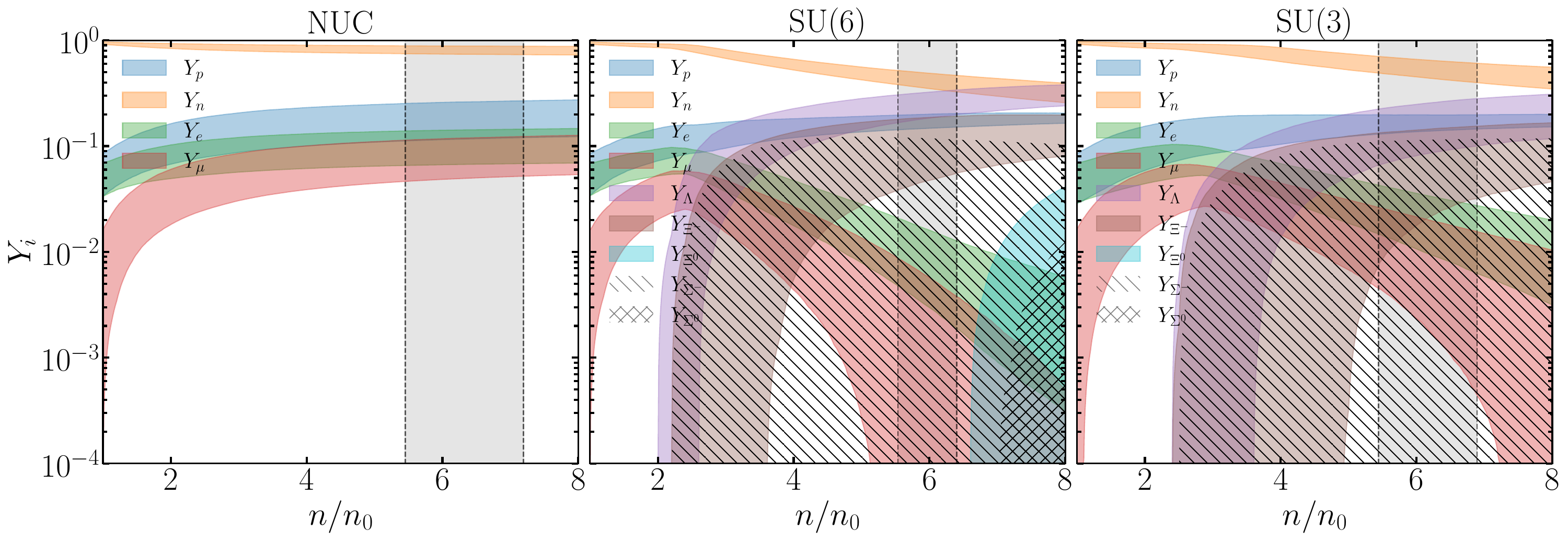}
 \caption{Particle fractions $Y_i$ as functions of the normalized baryon number density $n/n_0$. (Left panel) Nucleonic case; (middle panel) SU(6) symmetry scheme, and (right panel) SU(3) symmetry scheme. The vertical gray bands denote the 95\% CI of the central density for the maximum-mass star.}
    \label{fig:fraction}
\end{figure*}

    Figure~\ref{fig:fraction} illustrates the particle population fractions for the nucleonic, SU(6), and SU(3) symmetry schemes. A clear distinction emerges in the onset of hyperons, with species such as $\Lambda$ and $\Xi^{-}$ appearing at comparatively lower baryon densities in the SU(6) case than in the SU(3) framework. This behavior can be traced to the underlying meson-baryon coupling structure that governs the chemical potentials of the baryonic species in relativistic mean-field models. In general, the appearance of a new baryon is dictated by the condition that its chemical potential equals its in-medium effective energy, which depends on the interplay between scalar attraction and vector repulsion.

Within the SU(6) symmetry scheme, the vector meson couplings follow the QMIC relations that typically assign weaker couplings of hyperons to the $\omega$ meson, along with relatively modest contributions from the $\phi$ meson. As a result, the overall vector repulsion experienced by hyperons remains limited, allowing their chemical potentials to increase more rapidly with density. This facilitates an earlier fulfillment of the threshold condition, leading to the premature appearance of hyperons. In contrast, the SU(3) symmetry scheme introduces additional flexibility through symmetry-breaking parameters, namely $\alpha_v$ and $z_v$, and mixing angles ($\theta_v$), which generally enhance the repulsive vector interactions, particularly via stronger couplings to the $\omega$ and $\phi$ mesons. The increased repulsion shifts the effective energies of hyperons to higher values, thereby delaying their onset to larger densities. This is in agreement with the results reported by L\'opez \textit{et al.} \cite{2014PhRvC..89b5805L}, where it was shown that stronger vector repulsion suppresses the presence of hyperons at higher densities. In our case as well, the enhanced $\omega$ and $\phi$ meson couplings within the SU(3) framework lead to a delayed onset and subsequent reduction in hyperon fractions with increasing density.

\begin{figure*}
    \centering
    \includegraphics[width=1\linewidth]{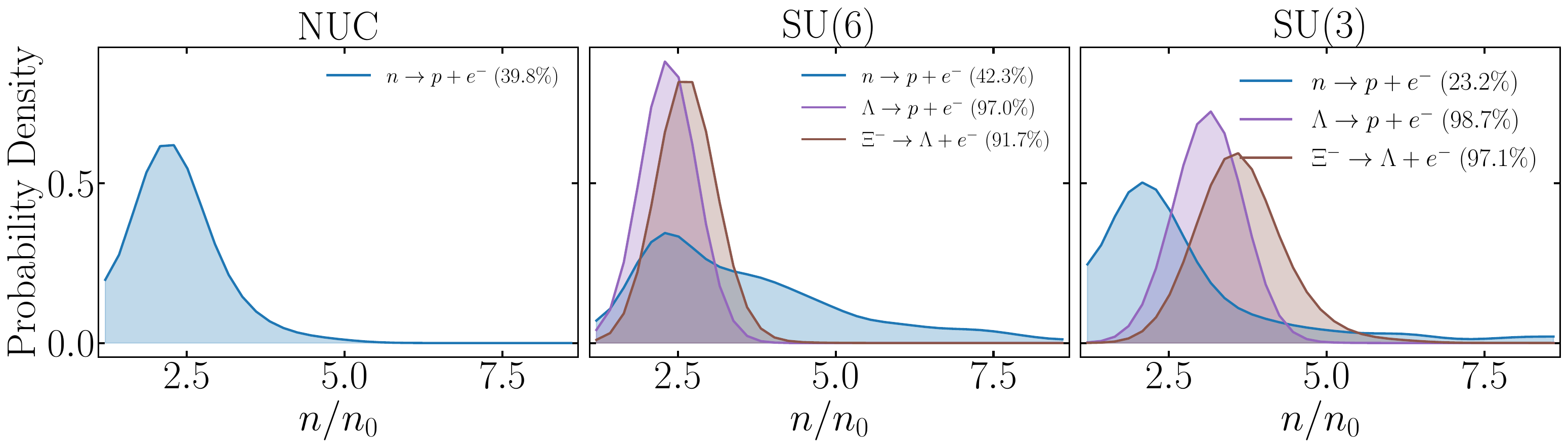}
    \caption{Probability density distributions of the threshold densities (in units of $n/n_0$) for the onset of direct Urca processes in NS matter for nucleonic, SU(6), and SU(3) scenarios.}
    \label{fig:durca}
\end{figure*}

This delayed onset of hyperons in the SU(3) case has important consequences for the stiffness of the EOS. Since the emergence of hyperons tends to soften the EOS by introducing additional degrees of freedom, postponing their appearance helps maintain a stiffer behavior over an extended density range. This is consistent with the comparatively larger values of the squared speed of sound obtained in the SU(3) scenario relative to SU(6). Owing to the exploration of the full parameter space, we also find that the $\Sigma^-$ hyperon can appear in matter in both SU(6) and SU(3) scenarios, with a large associated uncertainty. Although the $\Sigma^{0}$ and $\Xi^{0}$ hyperons appear in the SU(6) case, their onset occurs only beyond the density range reached in the maximum-mass configuration, and they do not appear in the SU(3) scenario within the density range considered here. This indicates that, within the present uncertainties of both the NS EOS and the hyperonic sector, the appearance of these neutral hyperons is disfavored, with SU(6) providing only a marginally more permissive threshold for their onset.

The direct Urca (DUrca) process plays a crucial role in governing the thermal evolution of NSs, as it provides an efficient mechanism for rapid neutrino emission from the stellar interior.

%In its simplest nucleonic form, the process involves beta decay and its inverse reaction \cite{643w-c2ly}, 
%\begin{equation}
%n \rightarrow p + e^- + \bar{\nu}_e
%\end{equation}
%and 
%\begin{equation}
%p + e^- \rightarrow n + \nu_e,
%\end{equation}
%which can proceed only when the proton fraction exceeds a critical threshold to satisfy momentum conservation. The inclusion of hyperons in dense matter opens additional DUrca channels, such as those involving $\Lambda$ and $\Xi^-$ baryons, which can significantly enhance the cooling rate
%\begin{equation}
%    \Lambda \rightarrow p + e^- + \bar{\nu}_e
%\end{equation}
%and
%\begin{equation}
%    \Xi^- \rightarrow \Lambda + e^- + \bar{\nu}_e.
%\end{equation}

A general DUrca process can be written as 
\begin{equation}
  B_1 \rightarrow B_2 + l + \bar{\nu}_l  
\end{equation} and its inverse 
\begin{equation}
    B_2 + l \rightarrow B_1 + \nu_l,
\end{equation}
where $B_1$ and $B_2$ denote baryons, $l$ represents a lepton (electron or muon), and $\nu_l$ ($\bar{\nu}_l$) is the associated neutrino (antineutrino). For the process to proceed, the phase-space momentum conservation condition, which is a triangular inequality, must be satisfied, which constrains the Fermi momenta as 
\begin{equation}
  |p_F^{B_2} - p_F^{l}| \leq p_F^{B_1} \leq p_F^{B_2} + p_F^{l}  
\end{equation}~\cite{1994A&A...290..458H}. Here, $p_F^{B_1}$, $p_F^{B_2}$, and $p_F^{l}$ are the Fermi momenta of the initial baryon, final baryon, and lepton, respectively. If this condition is violated, then the process is kinematically forbidden.

However, the onset of these processes is highly sensitive to the underlying EOS and the composition of matter, particularly the strength of baryon–-meson interactions that determine particle populations at high densities. Consequently, different symmetry schemes, such as SU(6) and SU(3), can lead to substantial variations in the threshold densities and the likelihood of DUrca processes. Figure~\ref{fig:durca} displays the probability density distributions of threshold densities for the initiation of direct Urca processes in NS matter under pure nucleonic, SU(6), and SU(3) frameworks. The panels show the distributions for the nucleonic process and hyperonic channels involving $\Lambda$ and $\Xi^-$ baryons. The percentages indicate the fraction of the parameter space for which each process is allowed. In the first panel, only the nucleonic process is present, as no hyperons are included. In this case, we also observe that the process is more favored around $2n_0$, whereas nearly 60\% of the EOSs in our posterior ensemble forbid the nucleonic DUrca process, suggesting that this behavior merits further scrutiny. In the SU(6) and SU(3) cases, hyperonic processes are also operative and occupy almost the entire fraction of the parameter space. In all three cases, the nucleonic DUrca process has more or less the same distribution percentage, thus implying that the inclusion of hyperons does not have much effect on the process. We also find that, in systems where hyperons are included, the hyperonic channels involving $\Lambda$ and $\Xi^-$ baryons dominate over a wide density range, indicating that their appearance is largely framework-independent once hyperons are present. We further identify the relevant DUrca processes, including both nucleonic and hyperonic channels. The onset of DUrca occurs at densities below $2n_0$ and up to $\approx 4n_0$, corresponding approximately to canonical $1.4\, M_\odot$ NSs, and is therefore significant for their thermal evolution and cooling. In the SU(6) scenario, we find a broad distribution of DUrca onset densities, reflecting the wider variation in the underlying parameter space.

\begin{figure}
    \centering
    \includegraphics[width=1\linewidth]{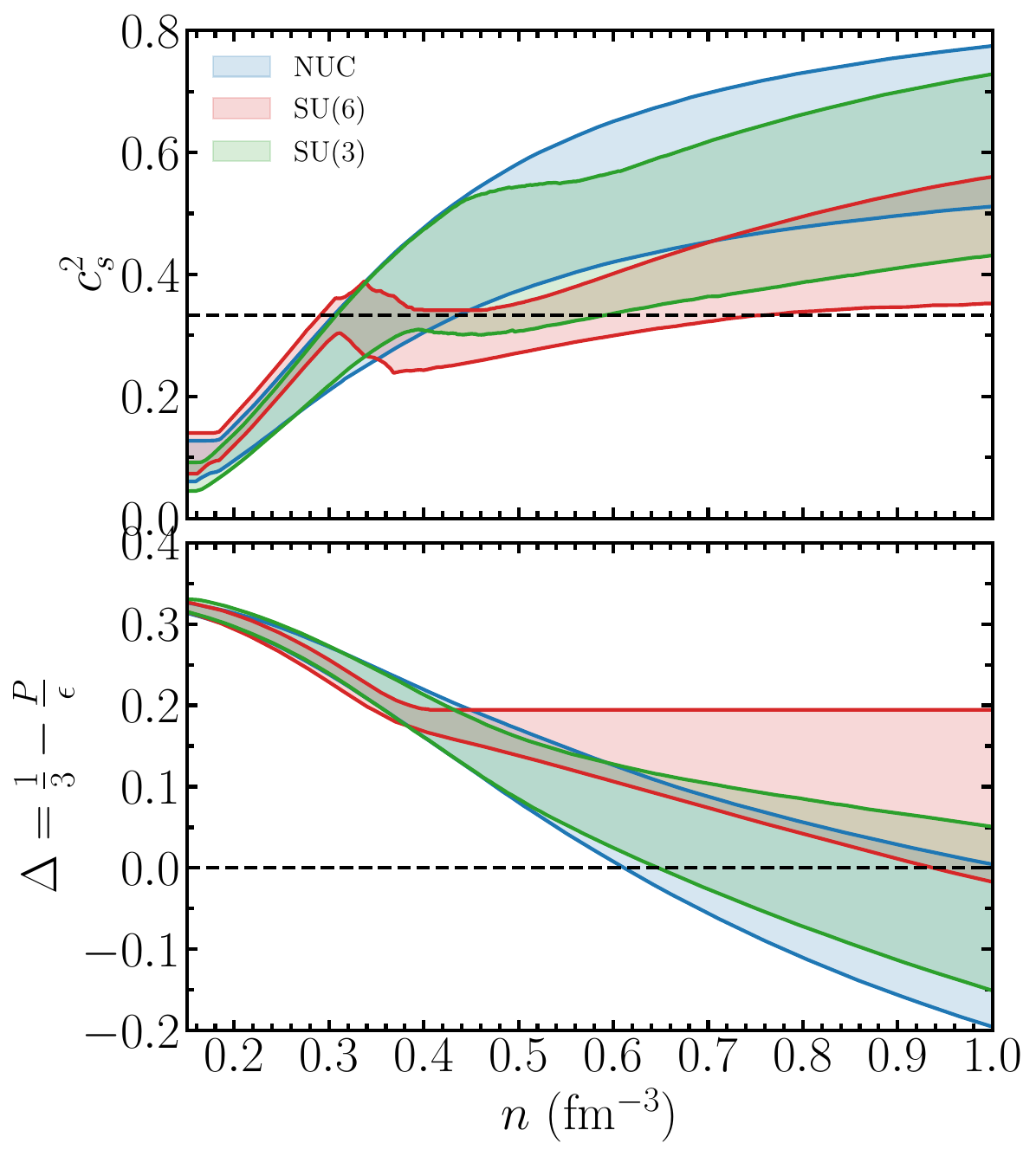}
    \caption{Top panel: Square of the speed of sound ($c_s^2$). Bottom panel: Trace anomaly ($\Delta$) as functions of baryonic density for the EOSs corresponding to the nucleonic, SU(6), and SU(3) symmetry schemes. The blue shaded band represents the nucleonic case, the red band corresponds to the SU(6) symmetry scheme, and the green band denotes the SU(3) symmetry scheme.}
    \label{fig:CsVSrho}
\end{figure}

%\begin{table*}[t]
%\centering
%\caption{Fitted parameters for the $f$-mode universal relations for different models. 
%Equation~(\ref{eq: freq}): $f = a \sqrt{M/R^3} + b$, and 
%Eq.~(\ref{eq: damping_time}): $R^4/(M^3 \tau_f) = a (M/R) + b$, 
%valid for $M/R > 0.15$ \cite{PhysRevC.99.045806}.}
%\begin{tabular}{lcccccc}
%\hline\hline
%Model & \multicolumn{3}{c}{Eq.~\eqref{eq: freq}} & \multicolumn{3}{c}{Eq.~\eqref{eq: damping_time}} \\
%\cline{2-4} \cline{5-7}
% & $a$ & $b$ & $R^2$ & $a$ & $b$ & $R^2$ \\
%\hline
%Nuc 
%& $34.581 \pm 0.007$ 
%& $0.6048 \pm 0.0003$ 
%& 0.983 
%& $-0.22042 \pm 0.00004$ 
%& $0.073614 \pm 0.000010$ 
%& 0.989 \\

%SU6 
%& $37.992 \pm 0.009$ 
%& $0.4814 \pm 0.0003$ 
%& 0.979 
%& $-0.22548 \pm 0.00004$ 
%& $0.074183 \pm 0.000008$ 
%& 0.992 \\

%SU3 
%& $34.005 \pm 0.006$ 
%& $0.6202 \pm 0.0002$ 
%& 0.980 
%& $-0.21881 \pm 0.00003$ 
%& $0.072909 \pm 0.000007$ 
%& 0.989 \\

%\hline\hline
%\end{tabular}
%\label{tab:fmode_fits}
%\end{table*}
%
 \begin{figure*}
    \centering
    \includegraphics[width=1\linewidth]{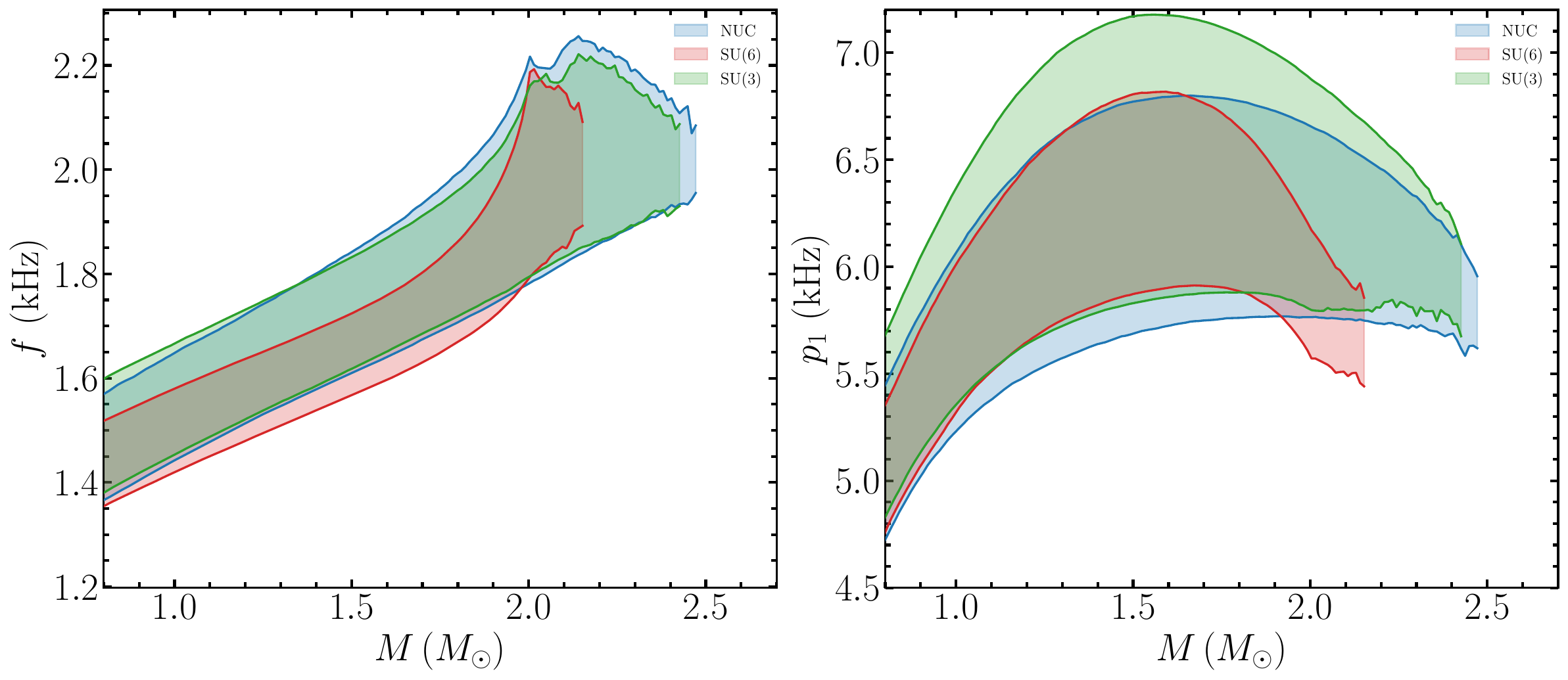}
    \caption{Left panel: The posterior probability distribution of f-mode oscillation frequency as a function of NS mass. Right panel: The posterior probability distribution of p1-mode oscillation frequency as a function of NS mass.}
    \label{fig:f_p1_mode}
\end{figure*}

Figure~\ref{fig:CsVSrho} illustrates the variation of the squared speed of sound, $c_s^2$, and trace anomaly, $\Delta$, as a function of baryon density $n$ for the three scenarios. At lower densities, all three cases exhibit a similar trend, with $c_s^2$ increasing steadily, indicating a stiffening of the EOS as density rises. However, as the density increases further, noticeable deviations emerge. The nucleonic case maintains a relatively higher value of $c_s^2$, reflecting a stiffer EOS due to the absence of additional degrees of freedom. In contrast, the inclusion of hyperons softens the EOS, leading to a reduction in $c_s^2$, with the SU(6) case showing the most pronounced softening. The SU(3) scenario, while still softer than the nucleonic case, exhibits comparatively higher values of $c_s^2$ than SU(6), suggesting that stronger repulsive interactions in the SU(3) framework partially counterbalance the softening effect of hyperons. The trace anomaly, defined as $\Delta = \frac{1}{3} - \frac{P}{E}$, is introduced as an indicator of conformality in NSs, and its role in dense matter physics is currently being actively explored \cite{Annala_2023, PhysRevC.107.025802, PhysRevD.108.043002, Chatterjee_2024}. In the conformal limit, $\Delta$ approaches zero. In all cases, $\Delta$ decreases monotonically with increasing density, indicating a gradual approach toward conformal behavior. For the purely nucleonic case, the decrease is more pronounced, with $\Delta$ crossing zero at intermediate densities and attaining increasingly negative values at higher densities. This behavior reflects a rapid stiffening of the EOS, leading to $P/E > 1/3$ beyond the conformal limit. In contrast, the SU(6) case exhibits a much smoother decline, with $\Delta$ remaining positive throughout the considered density range and approaching zero only asymptotically. This indicates that the system moves toward the conformal limit without overshooting it, consistent with the softening effect introduced by the early appearance of hyperons. The SU(3) case displays an intermediate behavior. While $\Delta$ follows a trend similar to the SU(6) case at lower densities, it decreases more rapidly at higher densities and eventually crosses zero, becoming mildly negative. This suggests a partial recovery of stiffness at high densities compared to SU(6), though less pronounced than in the nucleonic case. Overall, although all models exhibit a common trend toward conformality, their high-density behavior shows noticeable model dependence, reflecting the differing roles of hyperonic degrees of freedom and underlying symmetry assumptions.

NSs can sustain a rich spectrum of nonradial oscillation modes arising from perturbations in their fluid and spacetime structure, which serve as powerful probes of their internal composition and EOS. Among these, the fundamental ($f$) mode constitutes the lowest-order fluid mode and is primarily determined by the global stellar properties, such as mass, radius, and mean density \cite{PhysRevD.106.123002}. Its frequency exhibits a well-known scaling with the average density of the star, thereby providing a robust diagnostic of the underlying EOS \cite{PhysRevC.103.035810,643w-c2ly}. As a mode with no radial nodes \cite{PhysRevC.99.045806}, the $f$-mode is strongly coupled to the bulk motion of the fluid and typically dominates the oscillation spectrum, making it a promising source of gravitational-wave emission from perturbed NSs.

The posterior probability distribution of f-mode oscillation frequency as a function of NS mass is depicted in the left panel of Fig. \ref{fig:f_p1_mode}. These results highlight the universal character of the $f$-mode relations, indicating that it is not straightforward to infer the presence of hyperons solely from the behavior or shape of the mode, as their impact depends sensitively on the underlying symmetry assumptions adopted in the modeling. NSs support a spectrum of pressure ($p$) modes that arise from pressure gradients acting as the dominant restoring force. Among these, the first pressure mode ($p_1$) represents the lowest-order acoustic overtone and is characterized by the presence of one radial node in the eigenfunction. Unlike the $f$-mode, the $p_1$-mode is more sensitive to the local properties of the stellar interior, particularly the pressure profile and the stiffness of the EOS at higher densities \cite{PhysRevD.106.063005,1988ApJ...325..725M}. Its frequency generally lies above that of the $f$-mode and reflects the propagation of sound waves within the stellar medium, making it a useful probe of the internal structure and high-density behavior of NS matter.

\begin{figure*}
     \centering
    \includegraphics[width=1\linewidth]{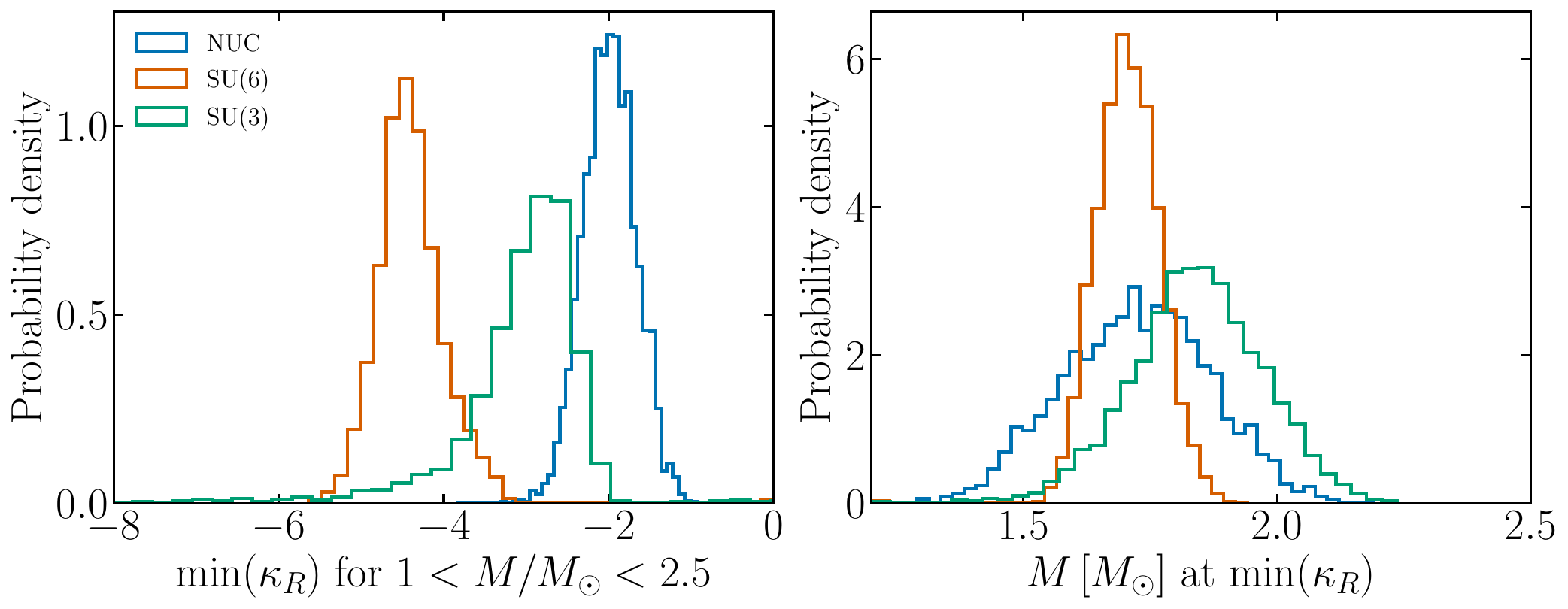}
    \caption{Left panel: Probability density distribution of the minimum value of curvature of the mass--radius relation $\kappa_R$ in the NS mass range $1 < M/M_\odot < 2.5$ for the nucleonic, SU(6), and SU(3) symmetry schemes. Right panel: Probability density distribution as a function of the stellar mass corresponding to the minimum value of $\kappa_R$.}
    \label{fig:signature}
\end{figure*}
Considerable effort has been devoted to identifying observable signatures of non-nucleonic degrees of freedom, such as hyperons, in NS interiors. The onset of hyperons is known to soften the EOS, typically leading to smaller radii and reduced maximum masses \cite{TOLOS2020103770, 2022atcc.book..153S, 2023PrPNP.13104041S, BURGIO2021103879, Schaffner-Bielich:2020psc}. However, translating this generic softening into a robust and model-independent observable has remained challenging. Earlier studies have primarily focused on global trends in the mass--radius ($M$--$R$) relation and tidal deformabilities, as well as on the slope $dR/dM$ as a potential diagnostic of EOS softening \cite{Lattimer_2001, PhysRevLett.121.161101}. Nevertheless, it has been shown that the slope alone does not provide a reliable discriminator, as nucleonic and hyperonic EOSs can exhibit significant overlap in their $dR/dM$ behavior across relevant mass ranges \cite{Weissenborn:2011kb, PhysRevC.94.035804}. More recently, attention has shifted toward higher-order structural properties of the $M$--$R$ relation, particularly its curvature, which captures the enhanced bending induced by the appearance of additional degrees of freedom.

In this work, we revisit one such diagnostic in the form of the curvature of the mass-radius relation, defined as \cite{bauswein2025}
\begin{equation}
\kappa_R = \frac{d^2R}{dM^2} \left[1 + \left(\frac{dR}{dM}\right)^2 \right]^{-3/2}.
\end{equation}
This quantity provides a direct measure of the geometric bending of the $R(M)$ curve and is more sensitive to EOS softening than the first derivative alone. As demonstrated in previous studies \cite{PhysRevD.110.123016}, hyperonic EOSs tend to exhibit significantly more negative values of $\kappa_R$ over intermediate mass ranges, reflecting the stronger softening induced by the onset of hyperons. In contrast, purely nucleonic models generally show a milder curvature in the same regime. The curvature thus offers a more discriminating probe of non-nucleonic degrees of freedom compared to the conventional slope $dR/dM$, which by itself is insufficient to uniquely identify their presence. The left panel of Fig.~\ref{fig:signature} shows the probability density distribution of the curvature of the mass–-radius relation, $\kappa_R$, evaluated over the mass range $1 < M/M_\odot < 2.5$ for different compositions. In the purely nucleonic scenario, the distribution of $\kappa_R$ remains largely above $-2.5$, indicating relatively mild curvature across the considered equations of state. In contrast, both the SU(6) and SU(3) hyperonic cases extend to lower values than the nucleonic one, indicating an overall shift toward enhanced curvature. However, the effect is markedly stronger in SU(6), whose distribution reaches significantly more negative $\kappa_{R}$ values, whereas the SU(3) peak remains much closer to the nucleonic case. This shows that the imprint of hyperons on the curvature is not universal but depends sensitively on the degree of EOS softening induced by the adopted symmetry scheme. Although Ref.~\cite{bauswein2025} identified a large negative curvature as a characteristic signature of hyperonic degrees of freedom, their conclusion was based on hyperonic models adopted from the literature, where the SU(6) symmetry scheme is commonly employed. The present results suggest that such an interpretation should therefore be viewed more cautiously and in a retrospective sense, since the strength of this signature is clearly model dependent.

%In contrast, for both SU(6) and SU(3) cases with hyperons, the distribution extends below this threshold, reflecting enhanced curvature, with the SU(6) case exhibiting the highest occurrence of EOSs attaining strongly negative curvature. This suggests that the presence of hyperons can leave a discernible imprint on the curvature, although its prominence depends sensitively on the underlying framework: stronger softening of the EOS leads to a more pronounced signature, whereas in comparatively stiffer cases this indicator becomes less effective.

The right panel of Fig.~\ref{fig:signature} displays the probability density distribution of the stellar mass corresponding to the minimum value of $\kappa_R$. A relatively sharp peak is observed in the SU(6) scenario; however, the overall mass range remains similar across all cases. Previous studies have suggested that the mass at which $\kappa_R$ attains its most negative value could serve as a signature of hyperon onset in NSs \cite{bauswein2025}. However, our results indicate that this mass range is largely unchanged across different compositions, implying that such a signature is not robust and must be interpreted with caution. In particular, its effectiveness depends sensitively on the softness of the underlying EOS. While the distributions for the nucleonic and SU(3) cases are relatively broad, the SU(6) case exhibits a pronounced peak, which can be attributed to the enhanced softening of the EOS at densities where hyperons appear, leading to a rapid variation in $\kappa_R$. This indicates that, although curvature may provide qualitative insights into the presence of hyperons, it cannot be regarded as a reliable or universal signature.

As mentioned earlier, a number of studies have demonstrated that the inclusion of additional repulsive mechanisms in hyperonic matter can significantly alter its high-density behavior. In particular, the incorporation of three-body forces involving hyperons has been shown to play a crucial role in reconciling hypernuclear properties with experimental observations while simultaneously modifying the composition and stiffness of dense matter. 
In the work \cite{Logoteta_2019}, the authors present that incorporating the chiral $NN\Lambda$ three‑body force into Brueckner‑Hartree‑Fock calculations both improves $\Lambda$ separation energies in heavy hypernuclei and stiffens the NS EOS enough to raise the predicted maximum mass to $\approx 2\, M_\odot$, thereby partially alleviating the hyperon‑puzzle. Such approaches effectively delay the onset of hyperons, suppress their abundance at higher densities, and lead to a stiffer EOS capable of supporting massive NSs. What is particularly noteworthy in these works is that, even after the inclusion of hyperonic degrees of freedom, the resulting matter exhibits bulk properties that closely resemble those of purely nucleonic matter. This is primarily achieved through the introduction of repulsive three-body interactions, which counterbalance the softening typically induced by hyperons.\\

%In a similar spirit, the present work employs the SU(3) symmetry framework to constrain the hyperon–meson couplings. Within this approach, we find that the resulting EOS and macroscopic properties of dense matter remain comparable to those of nucleonic matter, despite the presence of hyperons. Thus, our results reinforce the idea that appropriate theoretical constraints, whether through many-body forces or symmetry considerations, can effectively resolve the hyperon puzzle by restoring the stiffness of dense matter while maintaining consistency with known nuclear and astrophysical observations.

In a similar phenomenological spirit, the present work employs a generalized SU(3) symmetry framework to explore the allowed hyperon–-meson coupling space. Within this parametrization, we identify posterior-supported regions in which hyperonic matter remains compatible with the adopted nuclear and multimessenger constraints while producing bulk neutron-star properties close to those of nucleonic matter. This result should not be interpreted as a microscopic resolution of the hyperon puzzle. Rather, it demonstrates that the observational viability of hyperonic matter depends sensitively on the still poorly constrained interaction structure in the strange sector. Determining whether hyperons are present inside the neutron-star cores, and establishing the microscopic origin of the repulsion required at high density, therefore remains an open problem. Progress will require more precise hypernuclear measurements, improved constraints on hyperon–nucleon and hyperon-–hyperon interactions and hyperonic three-body forces, as well as increasingly informative multimessenger observations of neutron stars.

%\begin{figure}
%    \centering
%    \includegraphics[width=1\linewidth]{signature.pdf}
%    \caption{Variation of the curvature of the mass--radius relation $\kappa_R$ with maximum mass. The upper panel shows the nucleonic case, the middle panel corresponds to the SU(6) symmetry scheme, and the lower panel corresponds to the SU(3) symmetry scheme.}
%    \label{fig:K_RvsM}
%\end{figure}

\begingroup

\subsection{Bayesian model comparison}
\label{sec:bayesian_model_comparison}

\begin{table*}
\centering
\caption{Bayesian comparison of the nucleonic and hyperonic composition scenarios. The evidence difference is defined as $\Delta \ln Z_{ij} = \ln Z_i - \ln Z_j$, such that a positive value favors the model listed first. The Bayes factor is $B_{ij} = \exp(\Delta \ln Z_{ij})$.}
\label{tab:bayes_factors}
\begin{ruledtabular}
\begin{tabular}{lcccc}
Comparison & $\Delta \ln Z_{ij}$ & $B_{ij}$ & Approx.\ odds & Interpretation \\
\hline
Nucleonic vs SU(6) 
& $1.63 \pm 0.014$ 
& $5.10$ 
& $5.1:1$ 
& Positive evidence for nucleonic matter (weak-to-moderate)\\

Nucleonic vs\ SU(3) 
& $0.19 \pm 0.014$ 
& $1.21$ 
& $1.2:1$ 
& Inconclusive \\

SU(3) vs\ SU(6) 
& $1.44 \pm 0.014$ 
& $4.22$ 
& $4.2:1$ 
& Positive evidence for SU(3)  (weak-to-moderate)\\
\end{tabular}
\end{ruledtabular}
\end{table*}

The posterior distributions discussed above show that the present constraints affect the three composition scenarios differently. The nucleonic and SU(3) cases exhibit substantially overlapping posterior distributions for the nucleonic couplings, nuclear-matter properties, and the principal neutron-star observables.  In contrast, the restrictive SU(6) coupling prescription generally leads to earlier hyperon formation and a systematically softer equation of state.

To quantify the relative support for the three scenarios, we compare their Bayesian evidences. The nested-sampling analysis yields
\begin{align}
\nonumber
\ln Z_{\rm nuc} &= -57.06 \pm 0.01, \\
\ln Z_{\rm SU(6)} &= -58.69 \pm 0.01, \label{eq:logz_values}\\
\nonumber
\ln Z_{\rm SU(3)} &= -57.25 \pm 0.01,
\end{align}
where the quoted values denote the total evidence uncertainties reported by \textsc{UltraNest}. For each pairwise comparison, the uncertainty in the evidence difference is estimated as
$\sigma_{\Delta \ln Z} =
\sqrt{\sigma_{\ln Z_i}^{2}+\sigma_{\ln Z_j}^{2}}
\simeq 0.014$.

%The Bayes factor is defined as
%$B_{ij} = \exp\left(\Delta \ln Z_{ij}\right)$,
%where
%$\Delta \ln Z_{ij} = \ln Z_i-\ln Z_j$.

Table~\ref{tab:bayes_factors} summarizes the pairwise Bayesian evidence differences, corresponding Bayes factors, approximate posterior odds for equal prior model probabilities, and their qualitative interpretation for the three composition scenarios.
The nucleonic model has the largest evidence, but its difference relative to the SU(3) hyperonic model is negligible, with $\Delta \ln Z_{\rm nuc-SU(3)} = 0.19 \pm 0.014$ and $B_{\rm nuc,SU(3)} = 1.21$. Therefore, within the model space explored in this work, the present data do not provide meaningful statistical discrimination between a purely nucleonic composition and a hyperonic composition described by the more general SU(3) vector-coupling framework. In contrast, the nucleonic model is favored over the SU(6) framework by $\Delta \ln Z_{\rm nuc-SU(6)} = 1.63 \pm 0.014$, corresponding to $B_{\rm nuc,SU(6)} = 5.10$, while the SU(3) model is favored over SU(6) by $\Delta \ln Z_{\rm SU(3)-SU(6)} = 1.44 \pm 0.014$, corresponding to $B_{\rm SU(3),SU(6)} = 4.22$. These comparisons provide positive, but neither strong nor decisive, evidence against the restrictive SU(6) prescription relative to the nucleonic and SU(3) scenarios.

The Bayesian evidence should therefore not be interpreted as a direct detection or exclusion of hyperons. Rather, it indicates that the current combination of nuclear-matter, chiral-effective-field-theory, gravitational-wave, and NICER constraints does not strongly favor either the absence of hyperons or their presence in a sufficiently flexible interaction framework. The nearly equal evidence of the nucleonic and SU(3) scenarios shows that hyperonic matter can remain fully compatible with present constraints when the vector couplings are allowed to depart from the restrictive SU(6) relations. In the SU(3) case, the posterior distributions favor coupling combinations that enhance the repulsive vector interaction in the strange sector, delay the appearance of hyperons, and reduce their abundance over the density range relevant for most observed neutron stars. Consequently, the posterior EOS, mass–-radius relation, tidal deformability, and oscillation properties of the SU(3) model remain close to those of the purely nucleonic case.

By contrast, the SU(6) relations impose a more rigid hierarchy among the hyperon–meson couplings. Within the present DDRMF framework, this generally leads to weaker effective vector repulsion for hyperons, earlier hyperon onset, and a more pronounced softening of the EOS. The positive evidence against SU(6) therefore constrains this particular symmetry-based framework of the strange sector, rather than ruling out hyperonic degrees of freedom in neutron-star cores. In this sense, the available data are more informative about the allowed interaction structure of hyperonic matter than about the mere presence or absence of hyperons.

This result also has important implications for proposed observational signatures of hyperons. Many earlier studies of mass–-radius curvature, tidal deformability, maximum mass, direct-Urca thresholds, and oscillation properties have adopted SU(6)-motivated couplings or similarly restrictive hyperonic prescriptions. In such models, the early appearance of hyperons produces appreciable EOS softening and can generate pronounced structural signatures. Our posterior results show that these signatures are not generic consequences of hyperons themselves. When the broader SU(3) parameter space is considered, the enhanced vector repulsion can postpone hyperon onset and yield macroscopic observables that overlap substantially with the nucleonic posterior predictions. Therefore, a feature such as enhanced mass–-radius curvature, reduced maximum mass, or a distinctive oscillation pattern should be interpreted primarily as a signature of strong hyperon-induced softening rather than as an unambiguous signature of hyperon formation. This explains why the SU(6) case exhibits clearer deviations in the EOS, particle fractions, sound speed, curvature, and mode properties, whereas the SU(3) posterior remains much closer to the nucleonic scenario.

\endgroup

The principal outcome of the present analysis is therefore twofold. First, the close agreement between the nucleonic and SU(3) posterior predictions, together with their nearly equal Bayesian evidences, shows that current observations do not reliably distinguish a purely nucleonic composition from hyperonic matter when sufficient freedom is allowed in the strange-sector interaction. Second, the comparison with the SU(6) framework demonstrates that the same data can constrain restrictive assumptions about hyperon–-meson couplings. The present constraints thus provide conditional information about the interaction structure required for hyperons to remain compatible with neutron-star observations, rather than a direct detection or exclusion of hyperonic degrees of freedom.

\section{Summary}
\label{Summary}

\begingroup
    
We have carried out a Bayesian analysis of neutron-star matter within a density-dependent relativistic mean-field framework, comparing purely nucleonic matter with hyperonic matter described by the restrictive SU(6) prescription and by a generalized SU(3) vector sector with free parameters \(\alpha_v\) and \(z_v\). All three scenarios are confronted with the same set of nuclear, hypernuclear, chiral-EFT, heavy-ion, gravitational-wave, and NICER constraints, allowing their differences to be traced directly to the assumed composition and hyperon--meson interaction structure.

The restrictive SU(6) prescription leads to earlier hyperon onset, stronger softening of the equation of state, and reduced high-density pressure support. By contrast, the generalized SU(3) sector allows posterior-supported coupling combinations with stronger repulsion in the strange sector. These combinations delay the appearance of hyperons and reduce their abundance over the density range relevant for neutron-star interiors. As a result, the SU(3) posterior predictions for the equation of state, mass--radius relation, tidal deformability, and oscillation properties can remain close to those of purely nucleonic matter, whereas the SU(6) framework shows more pronounced deviations.

The Bayesian model comparison shows that the present combined constraints do not meaningfully distinguish purely nucleonic matter from hyperonic matter once the generalized SU(3) interaction freedom is admitted. In contrast, the restrictive SU(6) realization is comparatively less favored by the combined data. Thus, within the adopted DDRMF parametrization and priors, current information constrains restrictive assumptions about hyperon--meson couplings more effectively than it constrains the presence or absence of hyperons in neutron-star cores. The posterior also places substantial weight at values of \(\alpha_v\) below the SU(6) limit. The sensitivity analysis performed with different lower bounds on \(M_{\rm max}\) further shows that this result is not produced solely by the precise choice of the adopted \(2\,M_\odot\) threshold but arises from the combined likelihood, prior assumptions, and high-density stiffness requirements of the model.

Our results show that caution is required when interpreting proposed observational signatures of hyperons. In the restrictive SU(6) scenario, early hyperon onset and stronger equation-of-state softening can generate more visible changes in the mass--radius relation, tidal deformability, curvature of the mass--radius curve, and oscillation properties. However, these features become much weaker when the hyperon interaction sector is generalized through the SU(3) parametrization. In particular, the SU(3) posterior can overlap substantially with the nucleonic predictions for the principal macroscopic observables. Therefore, a reduced maximum mass, enhanced mass--radius curvature, or a modified oscillation spectrum should not be regarded as an unambiguous signature of hyperon formation. Such features are more directly associated with strong hyperon-induced softening under a particular interaction prescription.

The appearance of individual hyperon species remains model dependent within the posterior ensemble. In particular, the onset of \(\Sigma\) hyperons retains substantial uncertainty and depends sensitively on the allowed scalar and vector hyperon--meson couplings. The inferred direct-Urca threshold distributions identify cooling-relevant regions of the posterior parameter space. However, a quantitative comparison with neutron-star cooling observations requires dedicated thermal-evolution calculations that include neutrino emissivities, baryon pairing, and envelope physics.

The generalized SU(3) framework therefore provides phenomenologically viable hyperonic equations of state compatible with the present nuclear and multimessenger constraints, without requiring neutron-star observables to differ strongly from those of nucleonic matter. This should not be interpreted as a microscopic resolution of the hyperon puzzle, since the underlying hyperonic interactions remain only partially constrained and are not derived from first-principles many-body calculations. Improved hypernuclear information, microscopic constraints on hyperonic two- and three-body forces, and more precise multimessenger neutron-star observations will be required to reduce the remaining degeneracy between dense-matter composition and interaction structure.

\endgroup

\section*{Acknowledgements}
The authors thank the referee for the careful reading of the manuscript and for the constructive suggestions, which have helped to improve the clarity, scope, and presentation of the work.
The authors acknowledge the financial support from the Science and Engineering Research Board (SERB), Department of Science and Technology, Government of India, through project No. CRG/2022/000069. MS acknowledges the partial financial support from the DRDO through project No. DGTM/ERIP/GIA/24-25/010/005. 

% If you have a bibdatabase file and want BibTeX to generate the
%\bibliographystyle{elsarticle-num} 
\bibliography{references}
%\nocite{*}


\end{document}